\documentclass[preprintnumbers,showpacs,amsmath,amssymb,floatfix,9pt,prd,onecolumn,
superscriptaddress,nofootinbib]{revtex4}
\usepackage[utf8]{inputenc}
\usepackage{graphicx}
\usepackage{latexsym}
\usepackage{epsfig}
\usepackage{amssymb}
\begin{document}

\title{Charged strange star with Krori-Barua potential in $f(R,T)$ gravity admitting Chaplygin equation of state}

\author{Piyali Bhar}
\email{piyalibhar90@gmail.com
 } \affiliation{Department of
Mathematics,Government General Degree College, Singur, Hooghly, West Bengal 712 409,
India}

\begin{abstract}
In present paper a new compact star model in $f(R,T)$ gravity is obtained where $R$ and $T$ denote the Ricci
scalar and the trace of energy-momentum tensor $T_{\mu \nu}$ respectively. To develop the model we consider the spherically symmetric
space-time along with anisotropic fluid distribution in presence of electric field with $f(R,T)=R+2 \gamma T$ where $\gamma$ is a small positive constant. We have used the Chaplygin equation of state to explore
the stellar model. The field equations for $f(R,T)$ gravity have been solved by employing the Krori-Barua {\em ansatz} already reported in literature  [J. Phys. A, Math. Gen. 8:508,
1975]. The exterior spacetime is described by Reissner-Nordstr\"om line element for smooth matching at the boundary. It is
worthwhile to mention here that the values of all the constants involved with this model have been calculated for the strange stars 4U 1538-52 for different values of $\gamma$ with the help of
matching conditions. The acceptability of the model is discussed in details both analytically and graphically by studying the physical attributes of
matter density, pressures, anisotropy factor, stability etc. We have also obtained the numerical values in tabular form for central density, surface density, central
pressure and cental adiabatic index for different values of $\gamma$. The solutions of the field equations in Einstein gravity can be regained by simply putting $\gamma=0$ to our solution. Moreover, the proposed model is shown to be physically admissible
and corroborate with experimental observations on
strange star candidates such as 4U 1538-52.
\end{abstract}


\maketitle

\maketitle

\section{Introduction}

The study of compact object is a very interesting topic as it plays an important role to relate astrophysics, nuclear physics and particle physics. It is familiar that the neutron stars are built of neutron, on the other hand, strange stars are can be composed entirely of strange quark matter (SQM). Neutron stars are bounded by gravitational attraction and the strange stars are bounded by strong interactions as well as gravitational attractions. In 1915, Albert Einstein ignited the light among the scientific community by presenting one of the
greatest achievements of theoretical physics \cite{a1,a2,a3,a4}. The next year, Karl Schwarzschild \cite{karl} presented the first solution to the Einstein's field equations that describes the neighborhood
of a compact object that is spherically symmetric and static having vanishing pressure and density. A lot of pioneer works have been done till date in this direction \cite{j1,j2,j3,j4,j5,j6,j7}.

Bonnor proposed that if the matter present in the sphere carries certain modest electric charge density, a spherical body can remain in equilibrium under its own gravitation and electric repulsion, no internal
pressure is necessary \cite{bonnor}. Stettner \cite{set} studied the stability of a homogeneous distribution of matter containing a net surface and proved that a fluid sphere of uniform density with a modest surface charge is more stable than the same system without charge. According to De Felice et al.\cite{def} the gravitational collapse of a fluid sphere to a point singularity may be avoided in presence of large amounts of
electric charge during an accretion process onto a compact object.
Electrostatic repulsion due to the same electric charge along with
the pressure gradient counterbalance the gravitational attraction \cite{bek,gez}. The analysis of Raychaudhuri for charged
dust distributions showed that conditions for collapse and oscillation depend on the
ratio of matter density to charge density \cite{roya}. Di Prisco et al provided a full and comprehensive
analysis of charged, dissipative collapse \cite{dip}. A detailed analysis of
gravitational collapse in presence of charged medium was performed by Kouretsis and Tsagas
by highlighting the role of Raychaudhuri equation \cite{kou}. Maharaj and Takisa utilised an equation of state which is
quadratic relating the radial pressure to the energy density.


Alternative gravity is nowadays an extremely
important tool to address some persistent observational
issues, such as the dark sector of the universe. In current years, several
modified theories of gravity have been introduced but
a few theories like $f(R),\, f (T)$ and $f(R,T)$ have received
more attention than any other theories of gravity. The concept of $f(R,T)$ gravity was initially proposed by Harko {\em et al} \cite{harko11} by considering an
extension of standard general relativity in the year 2011, where the gravitational Lagrangian
is given by an arbitrary function of the Ricci scalar $R$ and of
the trace $T$ of the stress-energy tensor $T_{\mu \nu}$. They derived the field equations of the model
from a variational, Hilbert-Einstein type, principle and
covariant divergence of the stress-energy tensor is also
obtained. In that paper the author proved that, the covariant divergence of
the stress-energy tensor is nonzero and for this reason the motion of massive
test particles is nongeodesic, and an extra acceleration, due
to the coupling between matter and geometry, is always
present. It was also proposed by the author that the $f(R,T)$ gravity model depends on a source
term, representing the variation of the matter stress-energy
tensor with respect to the metric, and , its expression
can be obtained as a function of the matter
Lagrangian $L_m$ and hence for each choice of $L_m$
a specific set of field equations can be generated. However, Harko and his collaborators have constructed three possible models
by taking the functional $f(R, T)$, (i)~$f(R, T) = R + 2f(T)$,
(ii)~$f(R, T) = f(R) + f(T)$ and (iii)~$f(R, T) = f(R) + g(R)f(T)$, where $f(R),\,g(R)$
and $f(T)$ are some arbitrary functions of R and T.\par

Several astrophysical and cosmological model in the background of $f(R,T)$ gravity has been obtained by several authors.
Houndjo \cite{hou} studied a few distinguish cosmological
models in the context of $f (R,T )$ gravity to study a matter influenced
era of expanding Universe.
Yousaf et al. \cite{yousaf} discussed possibility development
about relativistic compact star over $f (R,T )$ gravity in
framework of Krori and Barua solution and two viable
$f (R,T )$ gravity models.
Moraes et al. \cite{moraes} studied hydrostatic equilibrium configurations
for the neutron stars and strange stars
by using Runge-Kutta 4th-order method to solve the TOV
equation in $f(R, T )$ gravity. Das et al. \cite{das1} studied
spherically symmetric and isotropic compact stellar system in $f(R, T)$ gravity by adopting
the Lie algebra with conformal Killing vectors and
also presented a model for Gravastar to avoid singularity in presence of $f(R,T)$ gravity \cite{das2}. Jamil et al. \cite{jamil} have reconstructed some cosmological models in this theory of
gravity using the functional form $f(R, T) = R^2 + f(T)$. Sahoo et al \cite{sahoo} obtained accelerating models in the
framework of $f(R, T)$ theory of gravity for an anisotropic Bianchi type III
(BIII) universe. Sharif and Zubair \cite{z1,z2} investigated perfect
uid distribution and mass less scalar
field for Bianchi type-I universe. Accelerating cosmological models have been constructed by Sahu et al. \cite{sahu} in  $f(R,T)$ modified
gravity theory at the backdrop of an anisotropic Bianchi type-III universe. Sahoo and his collaborators have reconstructed some $f(R, T)$
cosmological models for anisotropic universes \cite{s1,s2,s3,s4,s5}.

\par

Here in this work, we are interested to study
the charge effects within the framework of the $f(R,T)$ gravity
gravity in presence of Chaplygin equation of state (EoS).
We have arranged our present paper as follows. The basic field
equations of $f(R,T)$ gravity in presence of charge and the interior spacetime are given in Sec. \ref{sec2}. The field equations have been solved
by choosing suitable metric {\em ansatz} and a proper choice of an equation of state in Sec. \ref{sec3}. In order to fixed different constants we have matched our interior spacetime to the exterior Reissner-Nordstr\"om line element at the boundary outside the event horizon. The physical attributes of the model in modified gravity with the comparison of Einstein gravity have been shown in sec.~\ref{pa}. The next section describes about the mass radius relationship and the final section deals with some concluding remarks.
\section{Interior Spacetime and Basic field Equations}\label{sec2}
The Einstein Hilbert action for $f(R,T )$ gravity in presence of charged is given
by \cite{harko11},
\begin{eqnarray}\label{action}
S&=&\frac{1}{16 \pi}\int  f(R,T)\sqrt{-g} d^4 x + \int \mathcal{L}_m\sqrt{-g} d^4 x +\int \mathcal{L}_e\sqrt{-g} d^4 x,
\end{eqnarray}
where $f ( R,T )$ represents the general function of Ricci scalar $R$ and trace $T$ of the energy-momentum tensor $T_{\mu \nu}$, $\mathcal{L}_m$ and  $\mathcal{L}_e$ being the lagrangian matter density and Lagrangian for the electromagnetic field respectively with $g = det(g_{\mu \nu}$).\\
The field equations of the $f(R,T)$ gravity corresponding to action (\ref{action}) is given by,
\begin{eqnarray}\label{frt}
f_R R_{\mu \nu}-\frac{1}{2} g_{\mu \nu} f +(g_{\mu \nu }\Box-\nabla_{\mu}\nabla_{\nu})f_R &=& 8\pi (T_{\mu \nu}+E_{\mu \nu})\nonumber\\&&-f_T (T_{\mu \nu}+  \Theta_{\mu \nu}).
\end{eqnarray}
Where, $f=f(R,T)$, $f_R(R,T)=\frac{\partial f(R,T)}{\partial R},~f_T(R,T)=\frac{\partial f(R,T)}{\partial T}$. $\nabla_{\nu}$ represents the covariant derivative
associated with the Levi-Civita connection of $g_{\mu \nu}$, $\Theta_{\mu \nu}=g^{\alpha \beta}\frac{\delta T_{\alpha \beta}}{\delta g^{\mu \nu}}$ and
$\Box \equiv \frac{1}{\sqrt{-g}}\partial_{\mu}(\sqrt{-g}g^{\mu \nu}\partial_{\nu})$ represents the D'Alambert operator.\\
According to Landau and Lifshitz \cite{landau}, the stress-energy tensor of matter is defined as,
\begin{eqnarray}\label{tmu1}
T_{\mu \nu}&=&-\frac{2}{\sqrt{-g}}\frac{\delta \sqrt{-g}\mathcal{L}_m}{\delta \sqrt{g_{\mu \nu}}},
\end{eqnarray}
and its trace is given by $T=g^{\mu \nu}T_{\mu \nu}$. If the Lagrangian density $\mathcal{L}_m$ depends only on $g_{\mu \nu}$, not on its derivatives, eqn.(\ref{tmu1}) becomes,
\begin{eqnarray}
T_{\mu \nu}&=& g_{\mu \nu}\mathcal{L}_m-2\frac{\partial \mathcal{L}_m}{\partial g_{\mu \nu}}.
\end{eqnarray}
$E_{\mu \nu}$ is the electromagnetic energy-momentum tensor defined by,
\begin{eqnarray}
E_{\mu \nu}&=&\frac{1}{4 \pi}\left(F_{\mu}^{\alpha}F_{\nu \alpha}-\frac{1}{4}F^{\alpha \beta}F_{\alpha \beta} g_{\mu \nu}\right),
\end{eqnarray}
where, $F_{\mu \nu}$ denotes the antisymmetric
electromagnetic field strength tensor, defined by
\begin{eqnarray}
F_{\mu \nu}&=&\frac{\partial A_{\nu}}{\partial x^{\mu}}-\frac{\partial A_{\mu}}{\partial x^{\nu}},
\end{eqnarray}
satisfying the Maxwell equations,
\begin{eqnarray}
F^{\mu \nu}_{;\nu}=\frac{1}{\sqrt{-g}}\frac{\partial}{\partial x^{\nu}}(\sqrt{-g}F^{\mu \nu})&=&-4\pi j^{\mu},\label{ta}\\
F_{\mu\nu;\lambda}+F_{\nu \lambda;\mu}+F_{\lambda \mu;\nu}&=&0
\end{eqnarray}
where $A_{\nu}=(\phi(r),\,0,\,0,\,0)$ is the four-potential and  $j^{\mu}$ is the
four-current vector, defined by
\begin{eqnarray}
j^{\mu}&=&\frac{\rho_e}{\sqrt{g_{00}}}\frac{dx^{\mu}}{dx^0},
\end{eqnarray}
where $\rho_e$ denotes the proper charge density. The only non-vanishing components of
the electromagnetic field tensor are $F^{01}$ and $F^{10}$, and they are related by
$F^{01} = -F^{10}$, as for a static matter distribution the only non-zero component
of the four-current is $j^0$.
From Eq. (\ref{ta}) the expression for the electric field can be obtained as,
\begin{eqnarray}
F^{01}&=&-e^{\frac{\lambda+\nu}{2}}\frac{q(r)}{r^2},
\end{eqnarray}
here $q(r)$ represents the net charge inside a sphere of radius r can be obtained as,
\begin{eqnarray}
q(r)&=& 4\pi \int_0^r \rho_e e^{\frac{\lambda}{2}} r^2 dr.
\end{eqnarray}
Now the divergence of the stress-energy tensor $T_{\mu \nu}$ can be obtained by the taking covariant divergence of (\ref{frt}) (For details see ref \cite{harko11,hi,farri}) as,
\begin{eqnarray}\label{conservation}
\nabla^{\mu}T_{\mu \nu}&=&\frac{f_T(R,T)}{8\pi-f_T(R,T)}\Big[(T_{\mu \nu}+\Theta_{\mu \nu})\nabla^{\mu}\ln f_T(R,T)\nonumber\\&&+\nabla^{\mu}\Theta_{\mu \nu}-\frac{1}{2}g_{\mu \nu}\nabla^{\mu}T-\frac{8\pi}{f_T}\nabla^{\mu}E_{\mu \nu}\Big].
\end{eqnarray}

From eqn.(\ref{conservation}), we can check that $\nabla^{\mu}T_{\mu \nu}\neq 0$ if $f_T(R,T)\neq 0.$ So like Einstein gravity, the system will not be conserved. It can be noted that when $f(R,T)=f(R)$, from eqn. (\ref{frt}) we obtain
the field equations of $f(R)$ gravity.\par
The static and spherically symmetric line element in curvature coordinates $(t,\,r,\,\theta,\,\phi)$ is given by,
\begin{equation}\label{line}
ds^{2}=-e^{\nu}dt^{2}+e^{\lambda}dr^{2}+r^{2}d\Omega^{2},
\end{equation}
where $d\Omega^{2}\equiv \sin^{2}\theta d\phi^{2}+d\theta^{^2}$ and the metric co-efficients $\nu$ and $\lambda$ purely radial functions.
In the
present study we assume the compact star model with anisotropic
fluid. So, the stress-energy tensor of the
matter is given by,
\begin{eqnarray}
T_{\nu}^{\mu}=(\rho+p_r)u^{\mu}u_{\nu}-p_t g_{\nu}^{\mu}+(p_r-p_t)\eta^{\mu}\eta_{\nu},
\end{eqnarray}
where $\rho$ is the matter density, $p_r$ and $p_t$ are respectively the radial and transverse pressure in modified gravity, $u^{\mu}$ is the fluid four velocity satisfies the equations $u^{\mu}u_{\mu}=1$ and $u^{\mu}\nabla_{\nu}u_{\mu}=0$. Now the matter Lagrangian density, which
in the general case could be a function of both density and pressure, $L_m = L_m (\rho, p)$, or of
only one of the thermodynamic parameters, becomes an arbitrary function of the density of
the matter $\rho$ only, so that $L_m = L_m (\rho)$ \cite{har1}. For our present paper, the matter Lagrangian can be taken as $\mathcal{L}_m=\rho$ and the expression of $\Theta_{\mu \nu}=-2T_{\mu \nu}-pg_{\mu\nu}.$\\

In the relativistic structures, in order to discuss
the coupling effects of matter and curvature components in
$f (R, T )$ gravity, let us consider a
separable functional form given by,
\begin{eqnarray}
f (R, T ) = f_1(R)+f_2(T ),
\end{eqnarray}
$f_1(R)$ and $f_2(T)$ being arbitrary functions of $R$ and $T$ respectively. Several viable models
can be obtained in $ f (R, T )$ gravity by
choosing different forms of $f_1(R)$ along with linear combination
of $f_2(T)$. In our present model, we consider $f_1(R)=R$ and $f_2(T)=2\gamma T$, i.e., we choose
\begin{eqnarray}\label{e}
f(R,T)&=& R+2 \gamma T,
\end{eqnarray}
$\gamma$ is some small positive constant.
(which reduces to GR for $\gamma=0$ proposed by Harko et al. \cite{harko11}) to study the effects
of curvature-matter coupling. The term $2\gamma T$ induces time-dependent coupling (interaction) between curvature and
matter. It also corresponds to $\Lambda$CDM model with a time-dependent cosmological constant \cite{sharif1}.
Using (\ref{e}) into (\ref{frt})
The field equations in $f(R,T)$ gravity is given by,
\begin{eqnarray}
G_{\mu \nu}&=&8\pi (T_{\mu \nu}^{\text{eff}}+E_{\mu \nu}),
\end{eqnarray}
where $G_{\mu \nu}$ is the Einstein tensor and
\begin{eqnarray}
T_{\mu \nu}^{\text{eff}}&=& T_{\mu \nu}+\frac{\gamma}{8\pi}T g_{\mu \nu}+\frac{\gamma}{4\pi}(T_{\mu \nu}-\rho g_{\mu \nu}).
\end{eqnarray}
For the line element (\ref{line}), the field equations in modified gravity can be written as,
\begin{eqnarray}
8\pi\rho^{\text{eff}}+E^2&=&\frac{\lambda'}{r}e^{-\lambda}+\frac{1}{r^{2}}(1-e^{-\lambda}),\label{f1}\\
8 \pi p_r^{\text{eff}}-E^2&=& \frac{1}{r^{2}}(e^{-\lambda}-1)+\frac{\nu'}{r}e^{-\lambda},\label{f2} \\
8 \pi p_t^{\text{eff}}+E^2&=&\frac{1}{4}e^{-\lambda}\left[2\nu''+\nu'^2-\lambda'\nu'+\frac{2}{r}(\nu'-\lambda')\right]. \label{f3}
\end{eqnarray}

The quantity $q(r)$ actually
determines the electric field as,
\begin{eqnarray}
E(r)&=&\frac{q(r)}{r^2}.
\end{eqnarray}
where $\rho^{\text{eff}}$, $p_r^{\text{eff}}$ and $p_t^{\text{eff}}$ are respectively the density and pressures in Einstein Gravity where
\begin{eqnarray}
\rho^{\text{eff}}&=& \rho+\frac{\gamma}{8\pi}( \rho-p_r-2p_t),\label{r1}\\
p_r^{\text{eff}}&=& p_r+\frac{\gamma}{8\pi}(\rho+3p_r+2p_t),\label{r2}\\
p_t^{\text{eff}}&=& p_t+\frac{\gamma}{8\pi}(\rho+p_r+4p_t).\label{r3}
\end{eqnarray}
the prime denotes differentiation with respect to `r'. In next section we shall solve the eqns. (\ref{f1})-(\ref{f3}) to obtain the model of compact star.

\section{Choice of the metric potential and proposed model}\label{sec3}
In eqns. (\ref{f1})-(\ref{f3}), we have three eqns. with six unknowns. So we have to choose any three of them to make the system solvable. Now by our knowledge of algebra, we can choose it in $6C_3=20$ ways.\\
For our present model, we choose the coefficient of $g_{rr}$ and $g_{tt}$ as,
\begin{eqnarray}\label{elambda}
e^{\lambda}= e^{Ar^2}, e^{\nu}&=&e^{Br^2+C},
\end{eqnarray}
where $B,\,A$ are constants of dimension km$^{-2}$ and $C$ is dimensionless quantity.
Plugging (\ref{elambda}) into (\ref{f1})-(\ref{f3}), we obtain,
\begin{eqnarray}
\rho^{\text{eff}}+E^2&=&\frac{1 + e^{-A r^2} (-1 + 2 A r^2)}{r^2},\label{s1}\\
p_r^{\text{eff}}-E^2&=&\frac{-1 + e^{-A r^2} (1 + 2 B r^2)}{r^2},\label{s2}\\
p_t^{\text{eff}}+E^2&=&e^{-A r^2} \left(-A + 2 B + B (-A + B) r^2\right).\label{s3}
\end{eqnarray}
From eqns. (\ref{s1})-(\ref{s3}), one can note that if we choose the suitable expression of $E^2$, the expression for $\rho,\,p_r$ and $p_t$ can be obtained from eqns. (\ref{r1}) and (\ref{r2}).
Instead of choosing the expression for $E^2$, we consider chaplygin equation of state,
\begin{eqnarray}\label{s4}
p_r&=& \alpha \rho- \frac{\beta}{\rho^n},
\end{eqnarray}
where $\alpha,\,\beta$ and $n$ are positive constants. For our present model we took $n=1$ for the simplicity.


The Chaplygin gas model has a connection with string theory.  It can be obtained from the Nambu-Goto
action for a D-brane moving in a (D+2)-dimensional spacetime
in the light cone parametrization \cite{1,2,3}. This Chaplygin gas model has been supported different classes of observational tests such
as supernovae data \cite{22}, gravitational lensing \cite{23}, gamma ray bursts \cite{24}, cosmic mi-crowave background radiation \cite{25}. Later the Chaplygin equation of state has been
modified to a more generalized Chaplygin gas equation of state. The
generalised Chaplygin equation of state has been employed to model the compact objects. Mubasher et al. \cite{mubasher} constructed
a stationary, spherically symmetric, and spatially inhomogeneous
wormhole spacetime supported by a modified
Chaplygin gas. \\

Solving (\ref{s1})-(\ref{s3}), with the help of (\ref{s4}), we obtain,
\begin{eqnarray}
\rho^{\text{eff}}&=&\frac{1}{2 (1 + \alpha)}\left[\frac{(A + B) e^{-A r^2}}{4 \pi}+g_1(r)\right],\label{24}\\
p_r^{\text{eff}}&=&\frac{1}{2 (1 + \alpha)}\left[\frac{(1 + 2 \alpha) (A + B) e^{-A r^2}}{4 \pi}-g_1(r)\right],\label{25}
\\
p_t^{\text{eff}}&=&\frac{1}{8 \pi}\Big[e^{-A r^2} (-A + 2 B + B (-A + B) r^2) + \frac{-1 +
  e^{-A r^2} (1 + 2 B r^2)}{r^2}-\frac{4 \pi}{1 + \alpha}\Big\{\frac{f_1(r)}{ 4 \pi}-g_1(r)\Big\}\Big].\label{26}
\end{eqnarray}
the above equations provide the expression for matter density and pressures for Einstein gravity.\\
The expression of $E^2$ is obtained as,
\begin{eqnarray}\label{k4}
E^2&=&\frac{e^{-A r^2}}{(1 + \alpha) (3 \gamma + 4 \pi) r^2}\Big[-2 e^{A r^2} \gamma^2 r^2 g(r)+\gamma \Big\{2 \alpha (-1 + e^{A r^2}) + (4 A \alpha + B + 3 \alpha B) r^2\nonumber\\&& -
 12 e^{A r^2} \pi r^2 g(r)+(1 + \alpha) B (-A + B) r^4 + 2 (-1 + e^{A r^2} + A r^2)\Big\}-4 \pi \Big\{1 + (-A + B) r^2\nonumber\\&& - \alpha (-1 + e^{A r^2} + 2 A r^2)+e^{A r^2}\left(-1+4\pi r^2 g(r)\right)\Big\}\Big].
\end{eqnarray}
Now we want to solve the eqns. (\ref{r1})-(\ref{r3}), with the help of (\ref{24})-(\ref{26}) and obtain the expressions for matter density and pressures in modified gravity as,
\begin{eqnarray}
\rho&=&\frac{1}{2 (1 + \alpha)}\left[\frac{(A + B) e^{-A r^2}}{\gamma + 4 \pi}+g_2(r)\right], \label{l1}\\
p_r&=&\frac{1}{2 (1 + \alpha)}\left[\frac{(1 + 2 \alpha) (A + B) e^{-A r^2}}{\gamma + 4 \pi}-g_2(r)\right],\label{l2}\\
p_t&=&\frac{1}{6 \gamma + 8 \pi}\Big[e^{-A r^2} (-A + 2 B + B (-A + B) r^2) + \frac{-1 +
  e^{-A r^2} (1 + 2 B r^2)}{r^2}-\frac{2 (\gamma + 2 \pi}{1 + \alpha}\Big\{\frac{f_1(r)}{\gamma + 4 \pi}-g(r)\Big\}\nonumber\\&&-\frac{\gamma}{1+\alpha}\Big\{\frac{(A + B) e^{-A r^2}}{\gamma + 4 \pi}+g_2(r)\Big\}\Big].\label{l3}
\end{eqnarray}

The anisotropic factor $\Delta=p_t-p_r$ is obtained as,
\begin{eqnarray}
  \Delta &=&\frac{1}{6 \gamma + 8 \pi}\Big[e^{-A r^2} (-A + 2 B + B (-A + B) r^2) + \frac{-1 +
  e^{-A r^2} (1 + 2 B r^2)}{r^2}-\frac{2 (\gamma + 2 \pi}{1 + \alpha}\Big\{\frac{f_1(r)}{\gamma + 4 \pi}\nonumber\\&&-g_2(r)\Big\}-\frac{\gamma}{1+\alpha}\Big\{\frac{(A + B) e^{-A r^2}}{\gamma + 4 \pi}+g_2(r)\Big\}\Big]-\frac{1}{2 (1 + \alpha)}\left[\frac{(1 + 2 \alpha) (A + B) e^{-A r^2}}{\gamma + 4 \pi}-g_2(r)\right].
\end{eqnarray}

where, $g_1,\,g_2$ and $f_1$ are functions of $r$ are given as,
\begin{eqnarray*}
g_1(r)&=&\sqrt{4 (1 + \alpha) \beta + \frac{(A + B)^2 e^{-2 A r^2}}{16 \pi^2}},\\
g_2(r)&=&\sqrt{4 (1 + \alpha) \beta + \frac{(A + B)^2 e^{-2 A r^2}}{(\gamma + 4 \pi)^2}},\\
f_1(r)&=&(1 + 2 \alpha) (A + B) e^{-A r^2}.
\end{eqnarray*}

\begin{figure}[htbp]
    \centering
        \includegraphics[scale=.45]{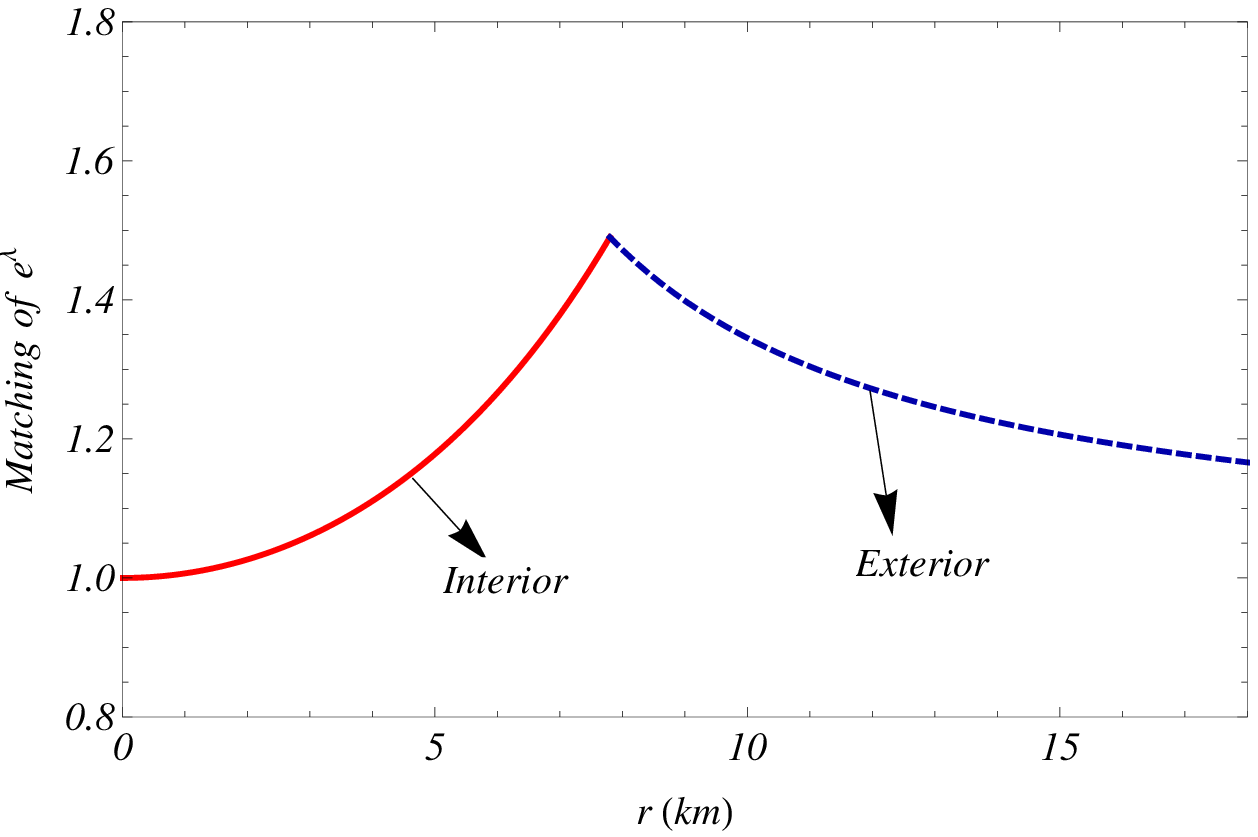}
        \includegraphics[scale=.45]{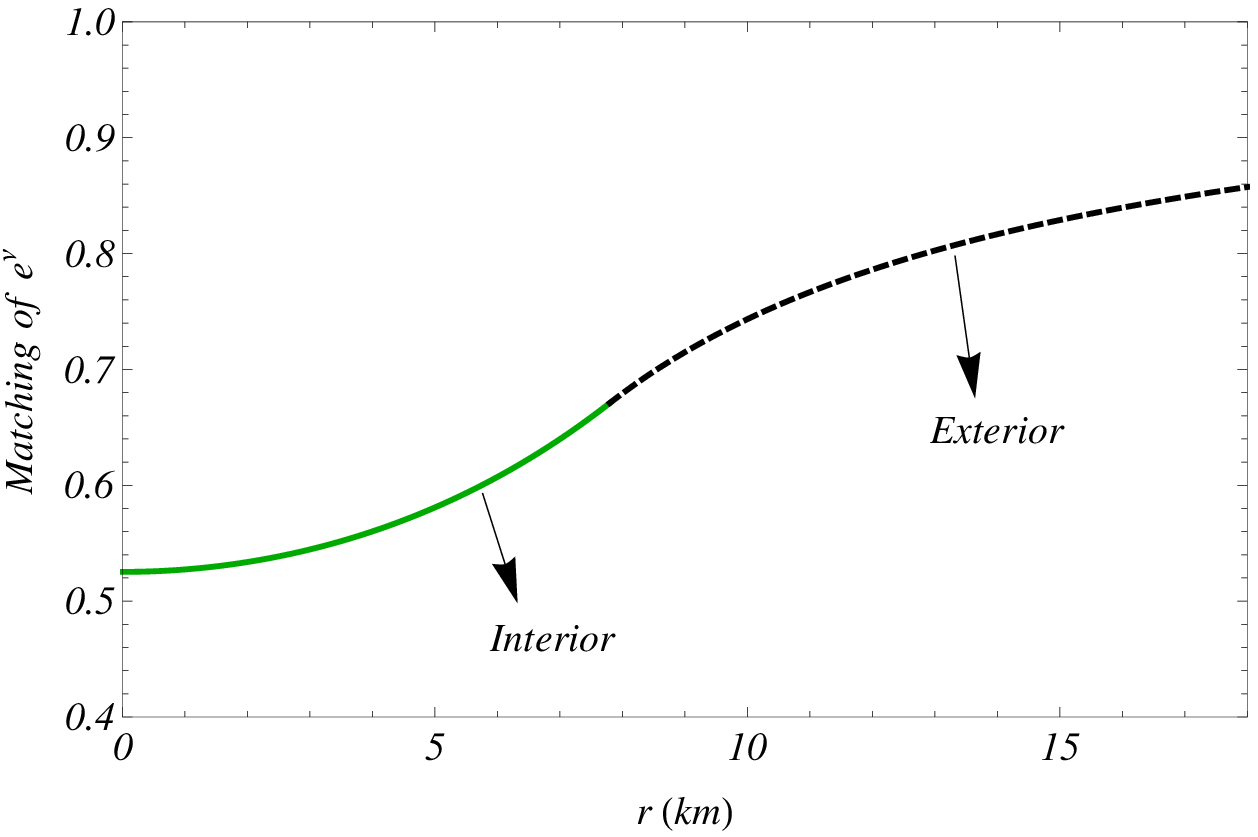}
       \caption{The matching condition of the metric potential $e^{\lambda}$ and $e^{\nu}$ are
shown against radius for the compact star 4U1538-52 by taking the values of
the constants $A,\,B$ and $Q$ mentioned in table~1.\label{metric}}
\end{figure}

\section{Boundary Condition}\label{bou}

In this section to fix the constants $A,\,B$ and $C$, we match our interior spacetime to the exterior spacetime outside the event horizon $r>M+\sqrt{M^{2}-Q^2}$, where, $Q$ is the total charge enclosed within the boundary $r=R$. The exterior space-time of the star will be described by the Reissner-Nordstr\"om metric \cite{rn1,rn21} given by
\begin{eqnarray}
ds^{2} &=& -\left(1 - \frac{2M}{r} + \frac {Q^2}{r^2}\right)dt^2 + \left(1 - \frac{2M}{r} + \frac {Q^2}{r^2}\right)^{-1}dr^2
\nonumber\\
&& + r^2(d\theta^2+\sin^2\theta d\phi^2), \label{eq22}
\end{eqnarray}
 Continuity of the metric coefficients $g_{tt}$, $g_{rr}$ and
$\frac{\partial g_{tt}}{\partial r}$ across the boundary surface $r= R$ between the interior and the exterior regions give the following set of relations:
\begin{eqnarray}
1 - \frac{2M}{R} + \frac {Q^2}{R^2} &=& e^{BR^2+C},\label{eq23}\\
1 - \frac{2M}{R} + \frac {Q^2}{R^2} &=& e^{-AR^2},\label{eq24}\\
\frac{M}{R^2} - \frac {Q^2}{R^3} &=& B Re^{BR^2+C}.\label{eq25}
\end{eqnarray}
Junevicus \cite{jun} obtained the expressions for $A,\,B,\,C$ from the
continuity of the first and second fundamental forms
across the surface of the charged fluid sphere in terms
of the dimensionless parameters $\frac{M}{R}$ and $\frac{|Q|}{R}$.\\
Eqs.~(\ref{eq23}) - (\ref{eq25}) determine the values of the constants $A$, $B$ and $C$ in terms of the total mass $M$, radius
$R$ and charge $Q$. By solving the above set of equations, we get
\begin{eqnarray}
A &=& - \frac{1}{R^2} \ln \left[ 1 - \frac{2M}{R} + \frac {Q^2}{R^2}
\right], \label{eq26}\\
B &=& \frac{1}{R^2} \left[\frac{M}{R} - \frac {Q^2}{R^2}\right] \left[1 - \frac{2M}{R} + \frac {Q^2}{R^2}
\right]^{-1},\label{eq27}\\
C &=&  \ln \left[ 1 - \frac{2M}{R} +
\frac {Q^2}{R^2} \right]- \frac{ \frac{M}{R} - \frac {Q^2}{R^2}}{
\left[ 1 - \frac{2M}{R} + \frac {Q^2}{R^2} \right]}. \label{eq28}
\end{eqnarray}\emph{}

We also impose the condition $E^{2}(r=0)=0$ which implies,
\begin{eqnarray}\label{bn1}
B (\gamma + 3 \alpha \gamma - 4 \pi) +
 2 A (2 + 3 \alpha) (\gamma + 2 \pi)=\nonumber\\
 2 (\gamma^2 + 6 \gamma \pi + 8 \pi^2)\sqrt{\frac{
  (A+B)^2 +
   4 (1 + \alpha) \beta (\gamma + 4 \pi)^2}{(\gamma + 4 \pi)^2}},\nonumber\\
\end{eqnarray}
and $p_r(r=R)=0$ gives,
\begin{equation}\label{bn2}
\alpha \rho_s^2=\beta,
\end{equation}
where $\rho_s$ is the surface density given by,
\begin{eqnarray}
\rho_s&=&\frac{1}{2 (1 + \alpha)}\left[\frac{(A + B) e^{-A R^2}}{\gamma + 4 \pi}+g_2(R)\right].
\end{eqnarray}
Solving (\ref{bn1}) and (\ref{bn2}) we get,
\begin{eqnarray}
\alpha&=&\frac{3 e^{2 A R^2} C_1\big((2 A-B)\gamma-4\pi(A-2B)\big)}{16 (A + B)^2 (\gamma + 2 \pi)^2-
  9 e^{2 A R^2} C_1^2},\label{b7}
  \\
  \beta&=&\frac{\alpha (A + B)^2 e^{-2 A R^2}}{(\gamma + 4 \pi)^2},\label{b8}
\end{eqnarray}
where, $C_1=(B +
    2 A )\gamma + 4A\pi. $ \par
Form eqns (\ref{eq26})-(\ref{eq28}) and (\ref{b7})-(\ref{b8}), it can be easily check that for a particular stellar model, by fixing the ratio of Mass to the radius and net charge to the radius one can obtain the values of $A,\,B,\,C,\,\alpha$ and $\beta$ for different choices of $\gamma$. It can also be checked that $\alpha$ and $\beta$ depends on $\gamma$ but $A,\,B$ and $C$ do not.

\begin{figure}[htbp]
    \centering
        \includegraphics[scale=.45]{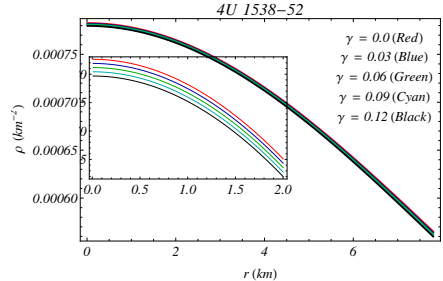}
       \caption{The matter density $\rho$ is plotted against $r$ inside the stellar interior for different values of $\gamma$ mentioned in the figure.\label{rho}}
\end{figure}

\section{Physical analysis of the present model}\label{pa}
In this section we shall check different physical attributes one by one both analytically and with the help of graphical representation.

\subsection{ Metric potential}

For our present model the metric potentials are chosen as,
\[e^{\nu}=e^{Br^2+C},~~~e^{\lambda}=e^{Ar^2},\]
We note that $e^{\nu}|_{r=0}=e^{C}>0$ and $e^{\lambda}|_{r=0}=1$, moreover,
\[\left(e^{\nu}\right)'=2 B e^{C + B r^2} r ,~~~~ \left(e^{\lambda}\right)'=2 A e^{A r^2} r.\]
Therefore, the form of the metric potential chosen here ensures
that the metric function is nonsingular, continuous, and well
behaved in the interior of the star. On a physical basis, this
is one of the desirable features for any well-behaved model.

\subsection{ Density and pressure }

The central density and central pressure for modified gravity is obtained as,
\begin{eqnarray}
\rho_c&=&\frac{1}{2(1+\alpha)}\left[\frac{A + B}{\gamma + 4 \pi} +D\right],\\
 p_c&=&\frac{1}{2(1+\alpha)}\Big[\frac{(1 + 2 \alpha) (A + B)}{\gamma + 4 \pi}-D\Big].
\end{eqnarray}
Where $D$ is a constant depending on $\gamma$ and its expression is given as,
\[D= \sqrt{
 4 (1 + \alpha) \beta + \frac{(A + B)^2}{(\gamma + 4 \pi)^2}}\]
The density and pressure gradients are obtained by taking the differentiation of the eqns.~(\ref{l1})-(\ref{l3}) with respect to r as,
\begin{eqnarray}
\frac{d\rho}{dr}&=&-\frac{A (A + B) e^{-2 A r^2} r}{(1 + \alpha) (\gamma + 4 \pi)^2}\bigg[e^{A r^2}(\gamma + 4 \pi)+h(r)\bigg],\label{k1}\\
      \frac{dp_r}{dr}&=&-\frac{A (A + B)r e^{-2 A r^2}}{(1 + \alpha) (\gamma + 4 \pi)^2}\bigg[(1 + 2 \alpha) e^{A r^2}(\gamma + 4 \pi)-h(r)\bigg],\label{k2}\\
      \frac{dp_t}{dr}&=&\frac{r}{6 \gamma + 8 \pi}\bigg[2 B (-A + B) e^{-A r^2}-\frac{2 A (A + B)e^{-2 A r^2} \gamma}{(1 + \alpha) (\gamma + 4 \pi)^2}\big(-e^{A r^2} (\gamma + 4 \pi) - h(r)\big)\nonumber\\
      &&+\frac{4 A (A + B) e^{-2 A r^2} (\gamma + 2 \pi)}{(1 + \alpha) (\gamma + 4 \pi)^2}\Big((1 + 2\alpha) e^{A r^2}\times(\gamma + 4 \pi)-h(r)\Big)-\frac{2 e^{-A r^2}(A - 2 B + 2 A B r^2)}{r^2}\nonumber\\&&+2 A e^{-A r^2} \Big(A - 2 B + (A - B) B r^2\Big) +
\frac{ 2 \Big(1 - e^{-A r^2} (1 + 2 B r^2)\Big)}{r^4}\bigg].\label{k3}
\end{eqnarray}
 \[\text{where,}~~~h(r)=\frac{A + B}{\sqrt{
   4 (1 + \alpha) \beta + \frac{(A + B)^2 e^{-2 A r^2}}{(\gamma +
      4 \pi)^2}}}\]
\begin{figure}[htbp]
    \centering
        \includegraphics[scale=.45]{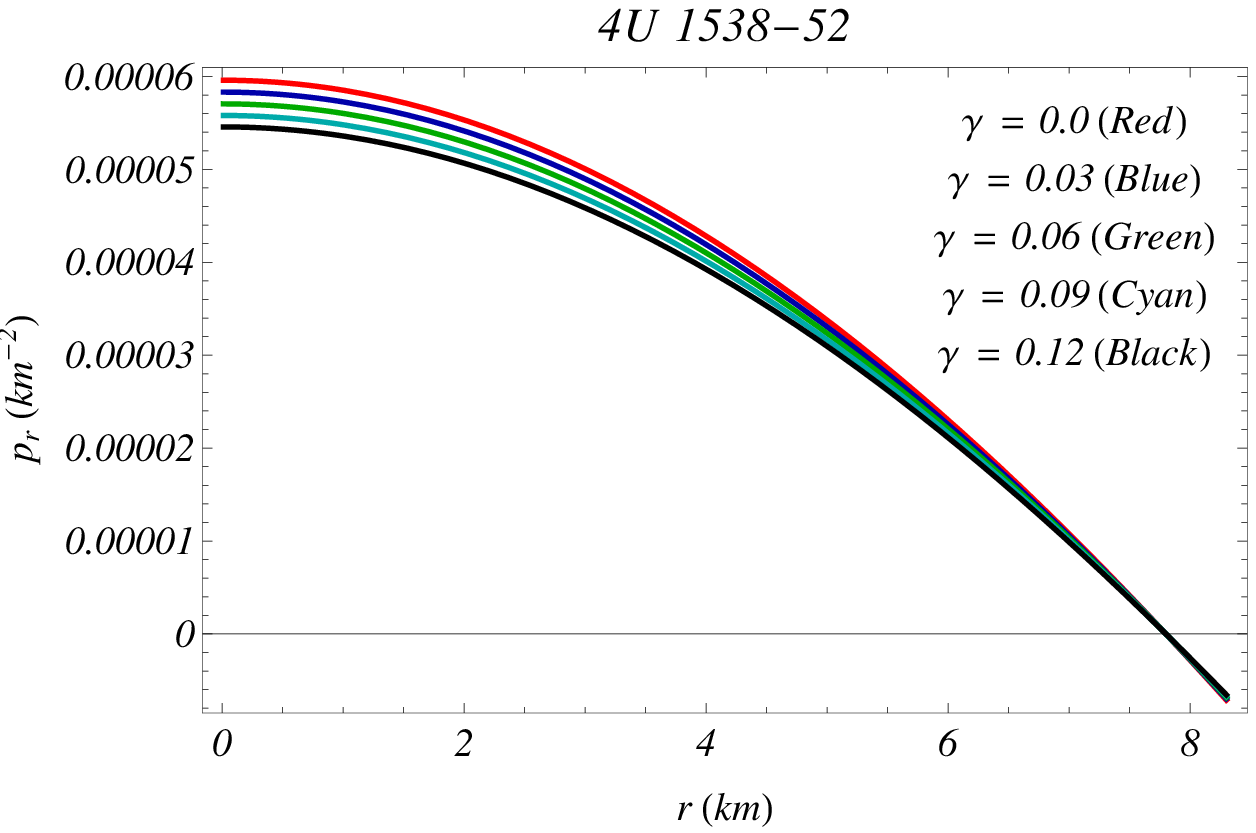}
        \includegraphics[scale=.45]{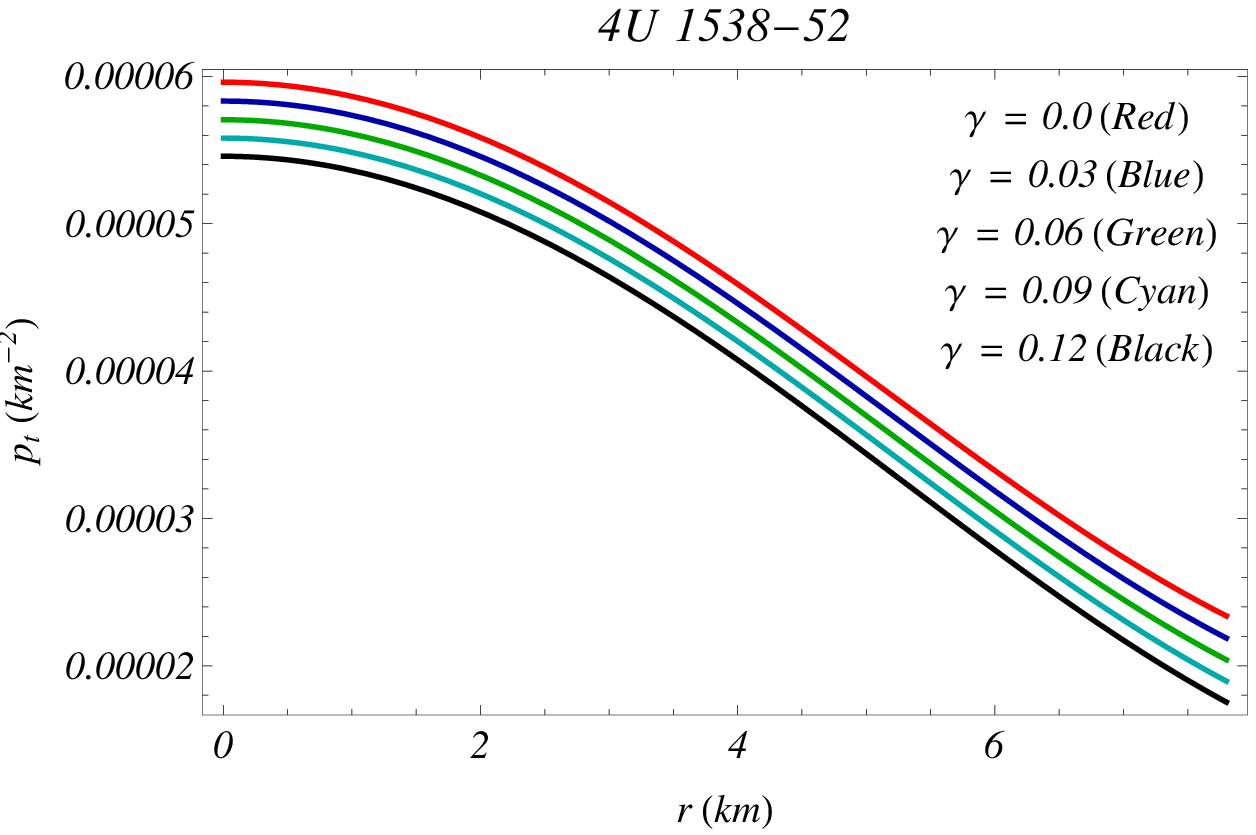}
        \includegraphics[scale=.45]{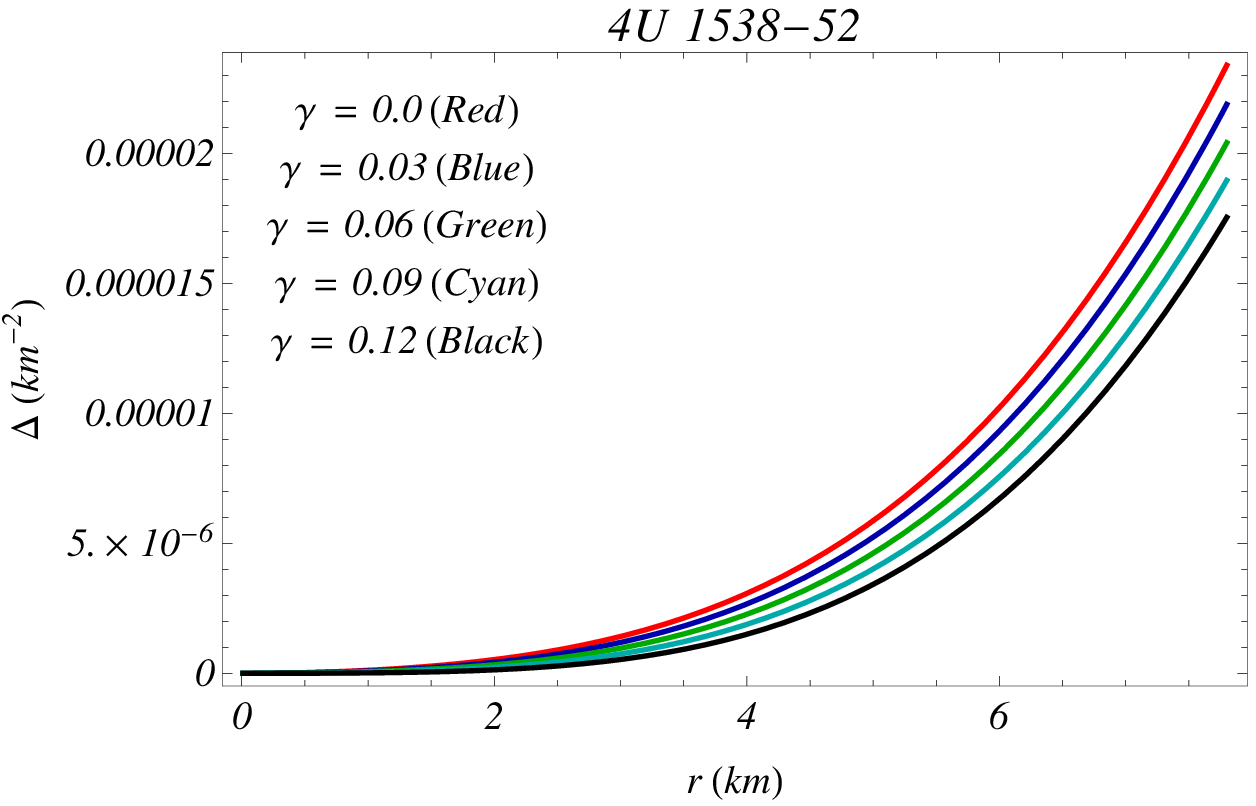}
       \caption{(left) Radial pressure $p_r$, (middle) transverse pressure $p_t$ and (right) anisotropic factor $\Delta$ are plotted against $r$ inside the stellar interior for different values of $\gamma$ mentioned in the figure. \label{pr}}
\end{figure}

\begin{figure}[htbp]
    \centering
        \includegraphics[scale=.45]{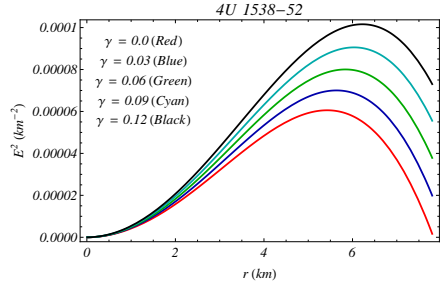}
       \caption{$E^2$ is shown against radius for different values of $\gamma$ mentioned in the figure.\label{ener}}
\end{figure}

The behavior of the metric potential $e^{\nu}$ and $e^{\lambda}$ have been shown in fig.~\ref{metric}. The matter density $\rho$ and both radial and transverse
pressures $p_r,\,p_t $ and the anisotropic factor $\Delta$ and the electric field $E^2$ versus the radial coordinate r for the compact star $4U 1538-52$
have been shown in figs.~\ref{rho}, \ref{pr} for the numerical values of the parameters mentioned in table~1. Both the anisotropic factor and electric field start
from zero at the center. The anisotropic factor increases towards the surface of the star, on the other hand, the electric field increase towards the surface of the star upto about $6$ km then it decreases towards boundary but at the surface of the star $E^2(R)>0,\,\Delta(R)>0$. The radial
pressure vanishes at the surface, neither the transverse pressure nor the matter density vanish there. The similar nature of the electric field $E^2$ was obtained by Maharaj et al \cite{maharaj10} to describe the models for quark stars with a linear equation of state.


\subsection{ Velocity of sound}

The radial and transverse velocity of sound $V_r$ and $V_t$ respectively for our present model is obtained as,
\[V_r=\sqrt{\frac{dp_r}{d\rho}},~~~~V_t=\sqrt{\frac{dp_t}{d\rho}}.\]

A model of compact star will be physically acceptable if both $V_r,\,V_t <1$ which is known as causality condition. On the other hand, according to Le Chatelier principle, the speed of sound must be positive. Combining the previous two cases, one can get, $0~<V_r,\,V_t<1$.\\
For our present model of compact object, the square of the radial and transverse speed of sound are obtained as,
\begin{eqnarray}
V_r^2&=&\frac{(1 + 2 \alpha) e^{A r^2} (\gamma + 4 \pi)-h(r)}{e^{A r^2} (\gamma + 4 \pi)+h(r)},\\
V_t^2&=&\frac{(1+\alpha)(\gamma+4\pi)^2 e^{2Ar^2}}{A(A+B)(6 \gamma + 8 \pi)}\bigg[2 B (A-B) e^{-A r^2}-\frac{2 A (A + B)e^{-2 A r^2} \gamma}{(1 + \alpha) (\gamma + 4 \pi)^2}\big(e^{A r^2} (\gamma + 4 \pi)+h(r)\big)\nonumber\\
      &&-\frac{4 A (A + B) e^{-2 A r^2} (\gamma + 2 \pi)}{(1 + \alpha) (\gamma + 4 \pi)^2}\Big((1 + 2\alpha) e^{A r^2}\times(\gamma + 4 \pi)-h(r)\Big)-\frac{2 e^{-A r^2}(A - 2 B + 2 A B r^2)}{r^2}\nonumber\\&&-2 A e^{-A r^2} \Big(A - 2 B + (A - B) B r^2\Big) -
\frac{ 2 \Big(1 - e^{-A r^2} (1 + 2 B r^2)\Big)}{r^4}\bigg].
\end{eqnarray}

Herrera and collaborators elaborately discussed
the concept of cracking for self-gravitating isotropic and anisotropic matter configurations in a series of lectures \cite{a10,a11,a12}. In 1992, L Herrera \cite{a10} introduced the concept of cracking (or overturning), that approach is useful to identifying potentially unstable anisotropic matter
configurations. He examined that inside the stellar interior, fluid elements, at both sides of the cracking point, are
accelerated with respect to each other. Later, Herrera along with his collaborators [7]
showed that even small deviations from local isotropy may lead to drastic changes in the
evolution of the system as compared with the purely locally isotropic case. Now it is easy to verify that,
\[\frac{\delta \Delta}{\delta \rho}~\sim~ \frac{\delta (p_t-p_r)}{\delta \rho}~\sim~ \frac{\delta p_t}{\delta \rho}- \frac{\delta p_r}{\delta \rho}~\sim~V_t^2-V_r^2. \]
\begin{figure}[htbp]
    \centering
        \includegraphics[scale=.45]{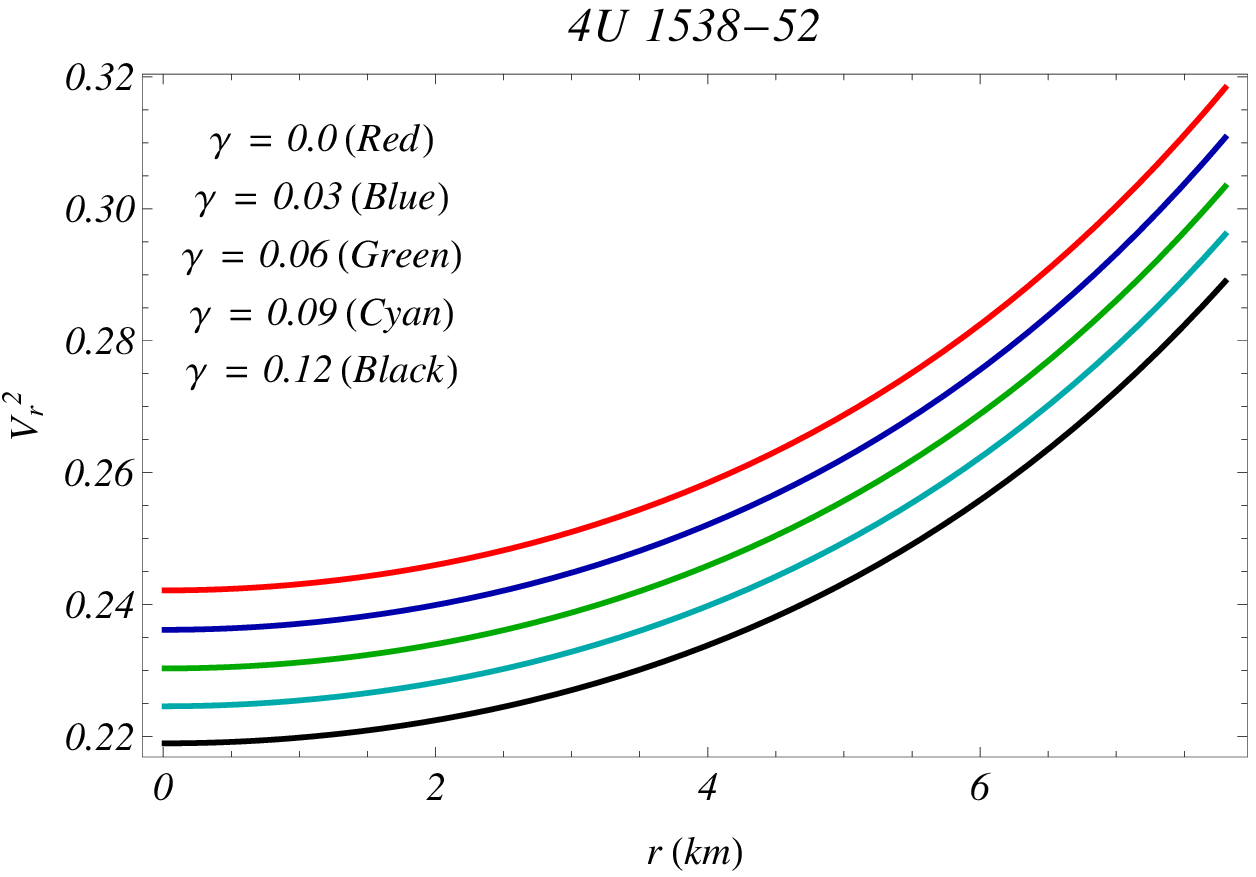}
        \includegraphics[scale=.45]{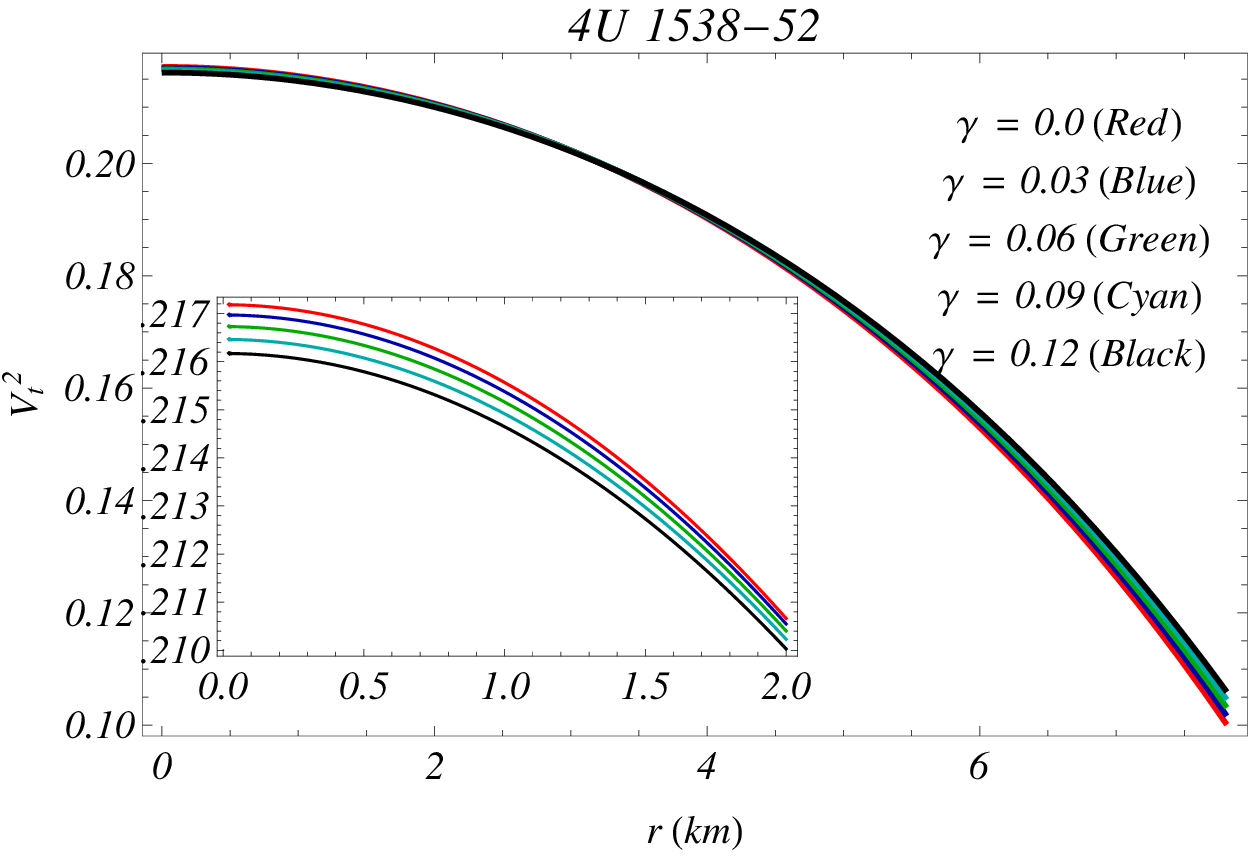}
        \includegraphics[scale=.45]{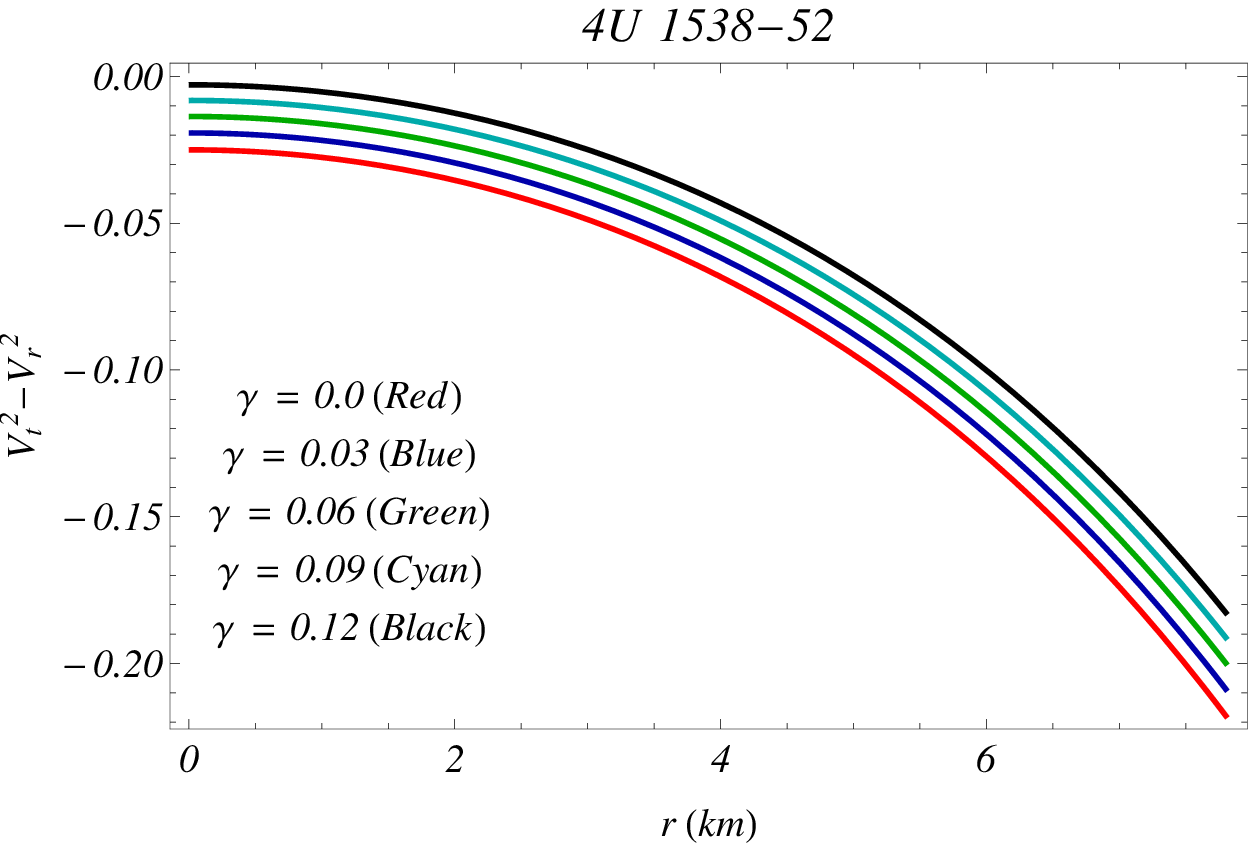}
       \caption{(left) Square of the radial sound velocity $V_r^2$, (middle) square of the transverse sound velocity $V_t^2$ and (right) the stability factor $V_t^2-V_r^2$ is plotted against r for the strange star candidate 4U 1538-52 by taking different values of $\gamma$.\label{sv}}
\end{figure}
Moreover, it is clear that for physically reasonable models, the magnitude of perturbations in anisotropy
should always be smaller than those in density since for physically acceptable stellar configuration,
\[|V_t^2-V_r^2|~\leq~1~\rightarrow~|\frac{\delta \Delta}{\delta \rho}|~\leq~1~\rightarrow~|\delta \Delta|\leq |\delta \rho|.\] This perturbations lead to potentially unstable models when $\frac{\delta\Delta}{\delta \rho}>0$.\\
To check the causality as well as the potentially stability criterion, we have shown the profiles of $V_r^2,\,V_t^2$ and $V_t^2-V_r^2$ in fig.~\ref{sv}. The figures show that $0~<V_r^2,\,V_t^2~<1$ holds everywhere inside the stellar interior and in the same time, $V_t^2-V_r^2~<0$, ensures the potential stability of the present model.

\subsection{Equilibrium under forces }
Using equations (\ref{f1})-(\ref{f3}), the generalized TOV equation for our present model in $f(R,T)$ gravity can be obtained as,
\begin{eqnarray}\label{con}
-\frac{\nu'}{2}(\rho+p_r)-\frac{dp_r}{dr}+\frac{2}{r}(p_t-p_r)+\frac{\gamma}{8\pi+2\gamma}(\rho'+p_r'+2p_t')\nonumber\\
+\frac{8\pi}{8\pi+2\gamma}\frac{q}{4\pi r^4}\frac{dq}{dr}=0,
\end{eqnarray}
In eqn.(\ref{con}), for $\gamma=0$ we regain the conservation equation in Einstein gravity with charged distribution. Now the above equation can be denoted by,
\begin{eqnarray}
F_g+F_h+F_a+F_e+F_m &=& 0.
\end{eqnarray}
\begin{figure}[htbp]
    \centering
        \includegraphics[scale=.45]{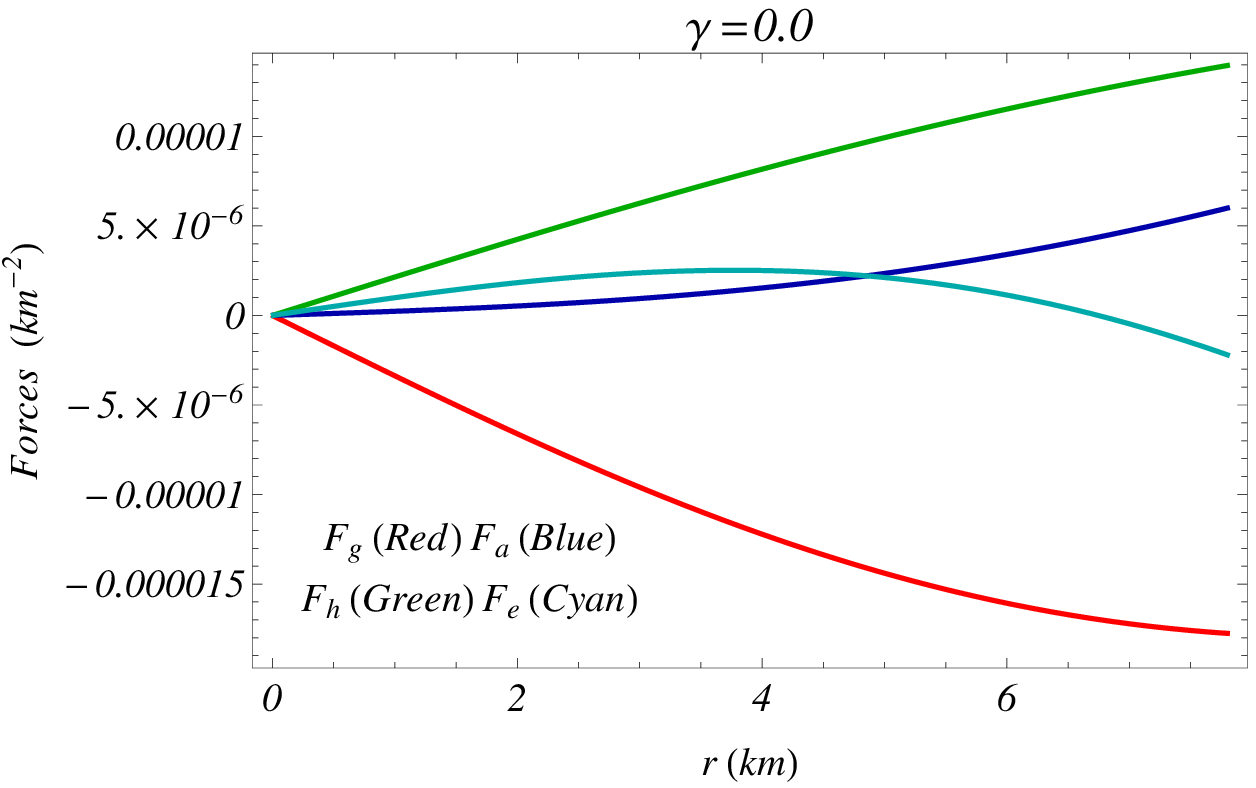}
        \includegraphics[scale=.45]{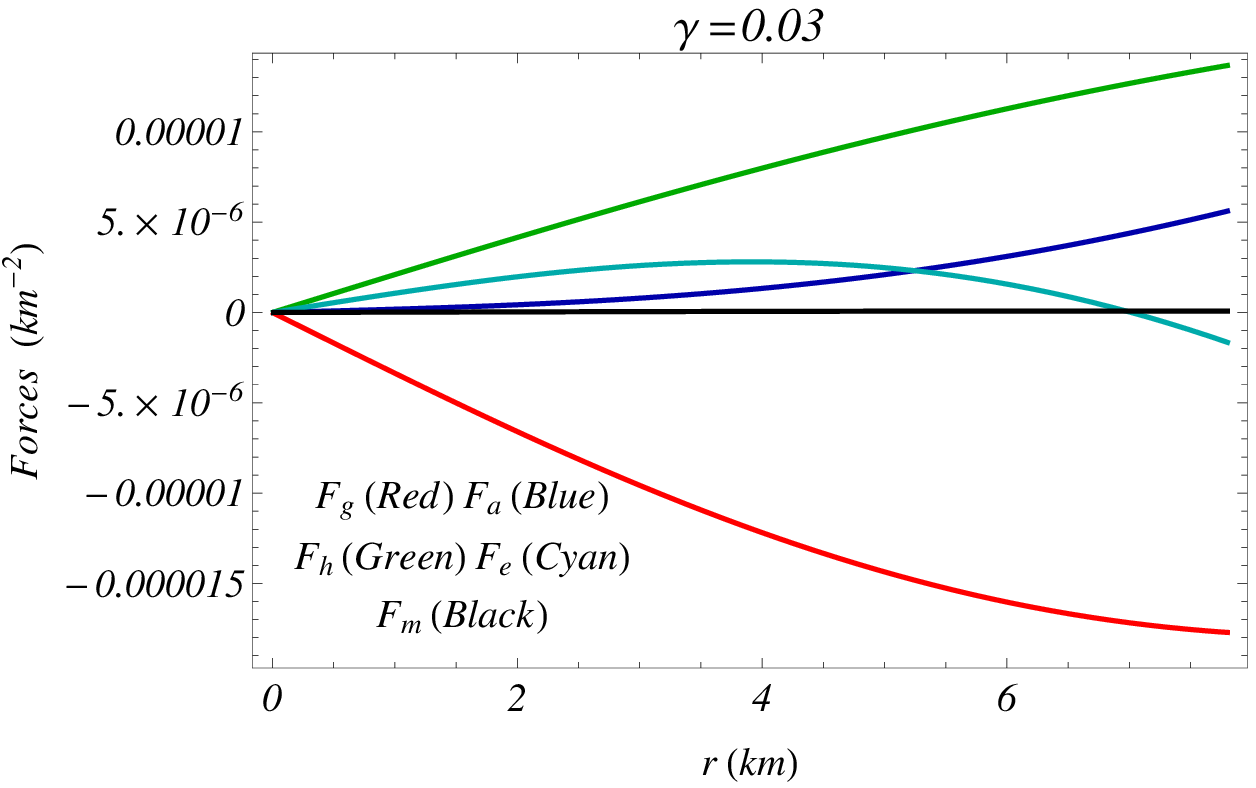}
        \includegraphics[scale=.45]{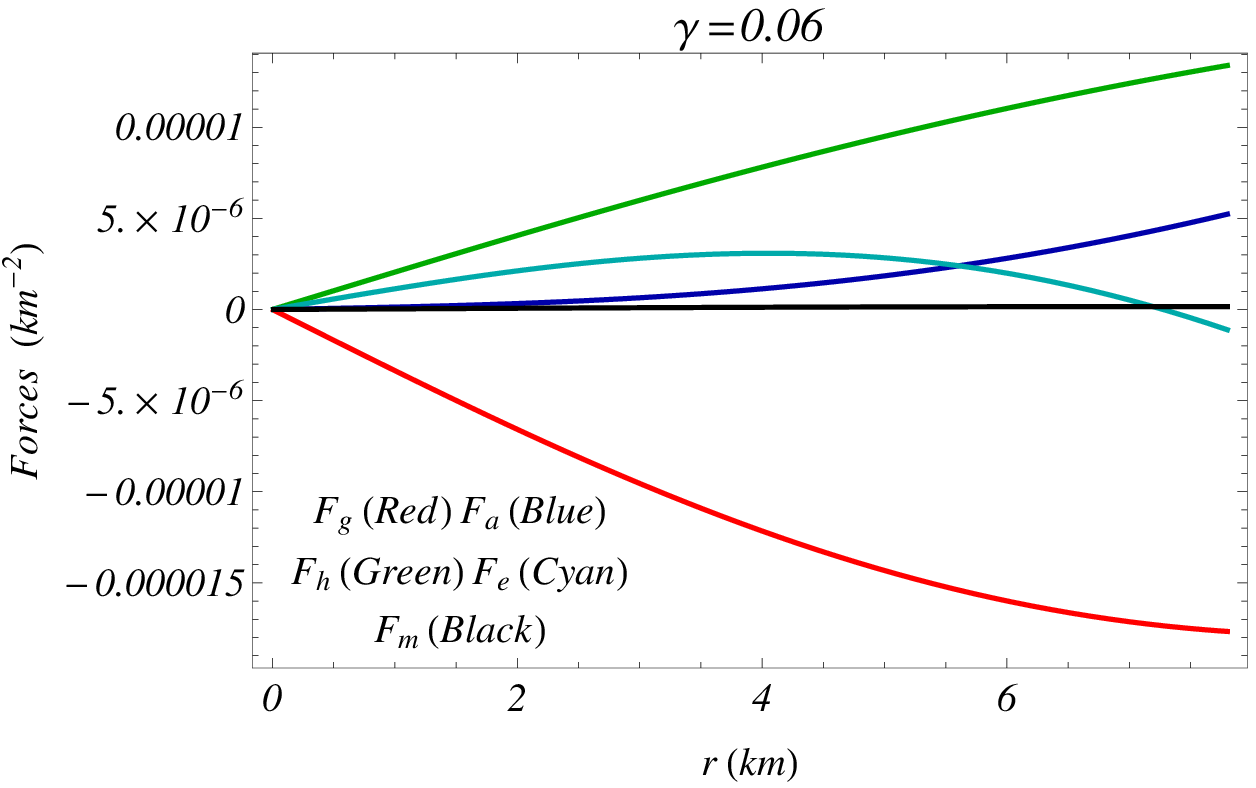}
        \includegraphics[scale=.45]{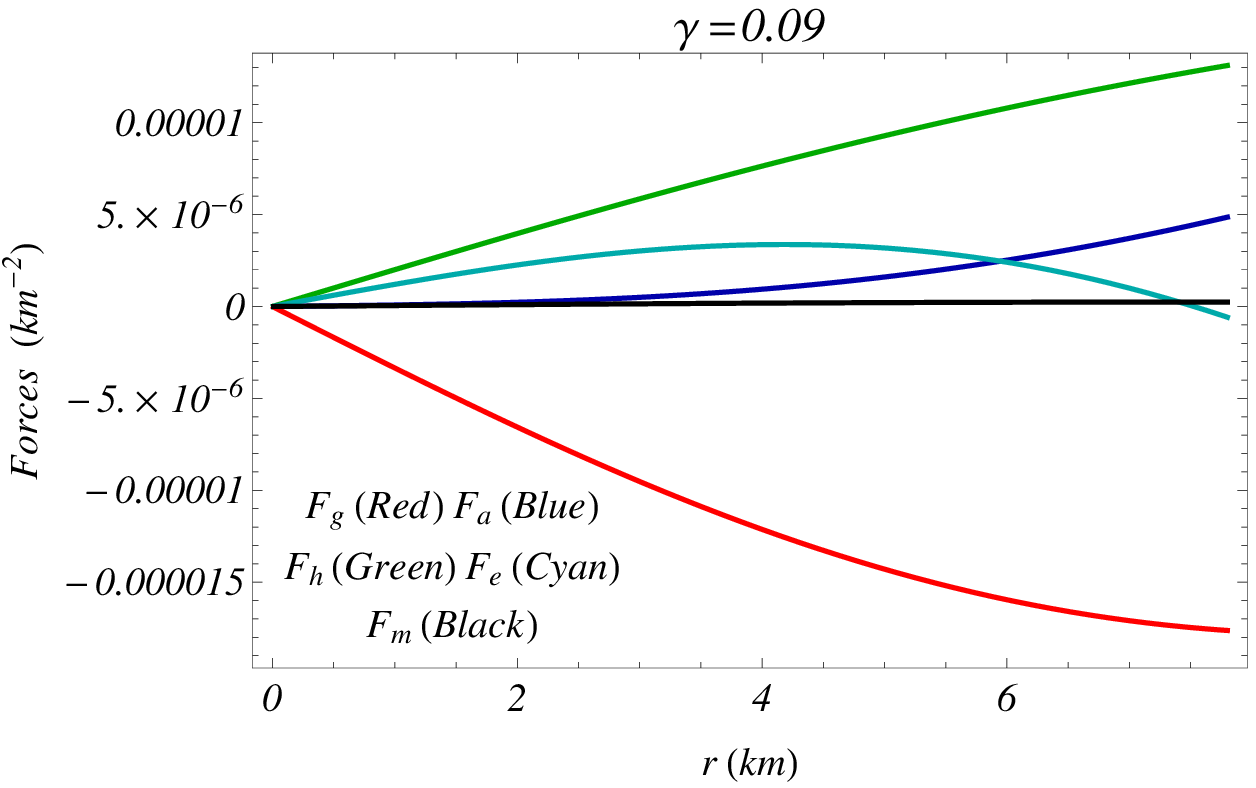}
        \includegraphics[scale=.45]{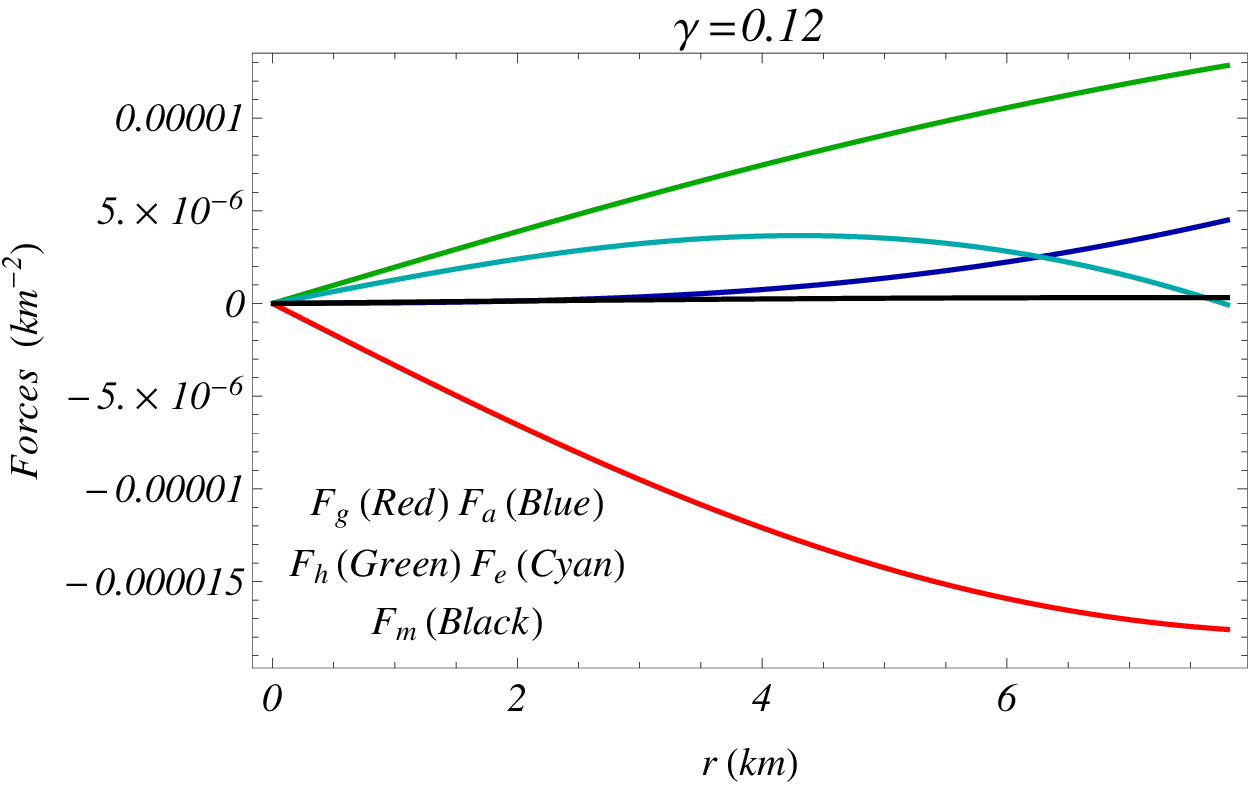}
       \caption{Different forces acting on the present model are plotted against r for the strange star candidate 4U 1538-52 by taking different values of $\gamma$.\label{tov}}
\end{figure}

Where, $F_g,\,F_h,\,F_a,\,F_e$ and $F_m$ respectively denote the gravitational force, hydrostatics force, anisotropic force, electric force and  force related to modified gravity and the expressions of the forces are given as,

\begin{eqnarray}
F_g&=&-\frac{\nu'}{2}(\rho+p_r)\nonumber\\
&=&-\frac{B (A + B) e^{-A r^2} r}{\gamma + 4 \pi},\\
F_h&=&-\frac{dp_r}{dr} \nonumber\\
&=&\frac{A (A + B)r e^{-2 A r^2}}{(1 + \alpha) (\gamma + 4 \pi)^2}\bigg[(1 + 2 \alpha) e^{A r^2}(\gamma + 4 \pi)-h(r)\bigg],\nonumber\\
F_a&=&\frac{2}{r}(p_t-p_r)=\frac{2}{r}\Delta \\
F_e &=&  \frac{8\pi}{8\pi+2\gamma}\frac{q}{4\pi r^4}\frac{dq}{dr} \nonumber \\
&=&\frac{1}{4\pi+\gamma}\left(\frac{2}{r}E^2+\frac{1}{2}\frac{d}{dr}(E^2)\right),\\
F_m&=& \frac{\gamma}{8\pi+2\gamma}(\rho'+p_r'+2p_t').
\end{eqnarray}
Where the expression for $E^2,\,\rho',\,p_r'$ and $p_t'$ are given in eqns. (\ref{k4}),(\ref{k1})-(\ref{k3}) respectively. The above mentioned forces for different values of $\gamma$ are shown in fig.~\ref{tov} and this figure verifies the equilibrium condition of the model of compact star.

\subsection{Relativistic Adiabatic index }
\begin{figure}[htbp]
    \centering
        \includegraphics[scale=.45]{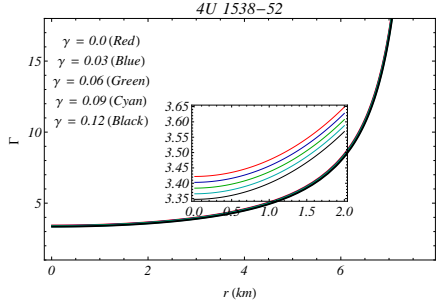}
       \caption{The relativistic adiabatic index $\Gamma$ has been plotted against r inside the stellar interior }\label{fgh1}
\end{figure}

Initially, Chandrasekhar \cite{255} did the pioneer work in this era to examine the stable/unstable
regions for spherical stars and explored the role of the adiabatic index. The adiabatic index $\Gamma$ for an isotropic fluid sphere was proposed
by Chan et al. \cite{chan10} as, $\Gamma=\frac{\rho+p}{p}\frac{dp}{d\rho}$. The expression for adiabatic index in case of pressure anisotropy changes as,
\begin{eqnarray}
\Gamma&=&\frac{\rho+p_r}{p_r}V_r^2,
\end{eqnarray}
The stability occurs if the adiabatic index is greater than $4/3$ as pointed out by Bondi \cite{bondi25}. For the complexity of the expression of expression of $\Gamma$ we will check this condition with the help of graphical representation. The profile of $\Gamma$ for different values of $\gamma$ is shown in fig.~\ref{fgh1}. We see that $\Gamma$ takes the values more than $4/3$ everywhere inside the fluid sphere.

\subsection{ Energy conditions }

There should be some restrictions among the model parameters $\rho,\,p_r,\,p_t$ and $E^2$ which play a crucial
role in understanding the nature of matter \cite{33}. For our anisotropic charged model, the four types of the energy conditions are satisfied if and only if the following inequalities hold everywhere inside the fluid sphere.
\begin{itemize}
\item Weak Energy Condition (WEC):~$\rho+p_r \geq 0,~\rho + p_t +\frac{E^2}{4\pi} \geq 0,~ \rho + \frac{E^2}{8\pi} \geq 0,$
\item Strong Energy Condition {SEC:~}$\rho+p_r \geq 0,~\rho + p_t +\frac{E^2}{4\pi} \geq 0,\nonumber\\ \rho+ p_r +2 p_t+ \frac{E^2}{4\pi} \geq 0,$
\item Dominant energy condition {DEC:~}$\rho-p_r +\frac{E^2}{4 \pi}\geq 0,~\rho - p_t \geq 0,~ \rho + \frac{E^2}{8\pi} \geq 0.$
\item Null energy condition (NEC):~$\rho+p_r \geq 0,~\rho + p_t +\frac{E^2}{4\pi} \geq 0,$
\end{itemize}

\begin{figure}[htbp]
    \centering
        \includegraphics[scale=.45]{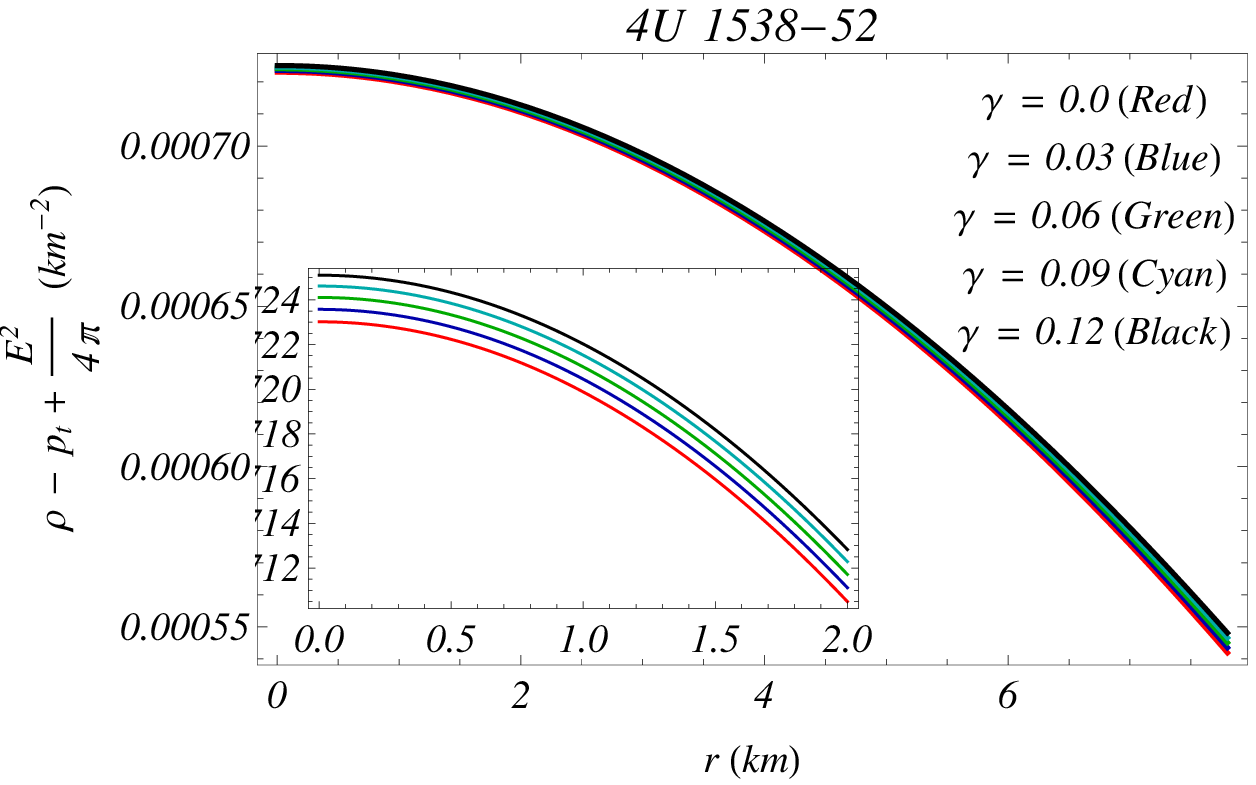}
        \includegraphics[scale=.45]{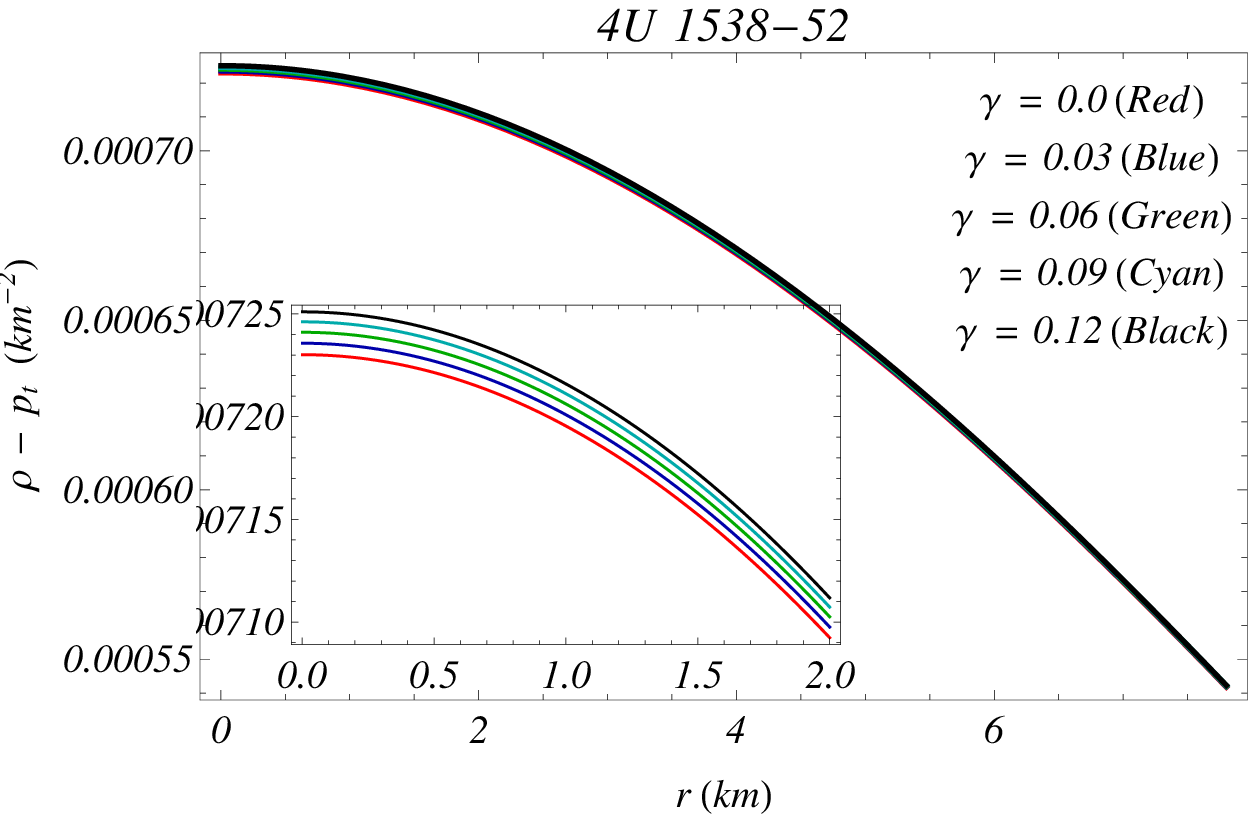}
        \includegraphics[scale=.45]{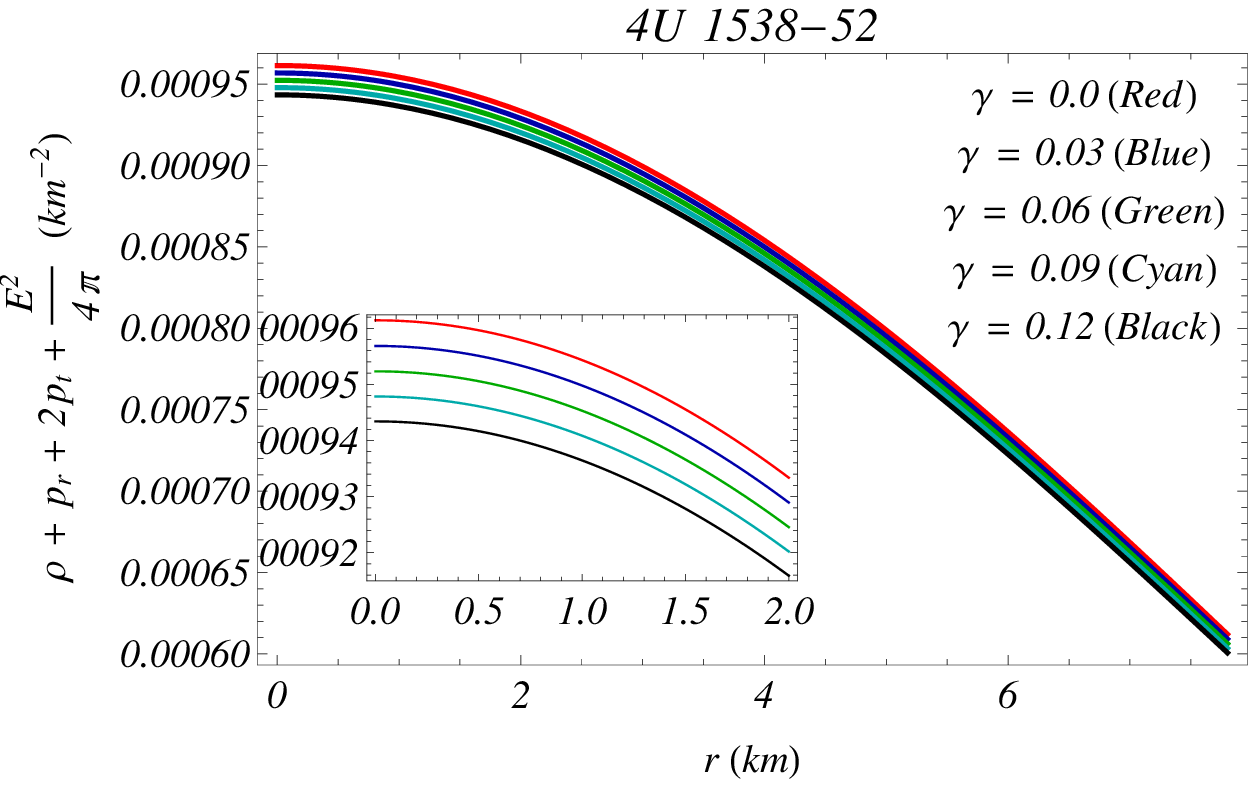}
        \includegraphics[scale=.45]{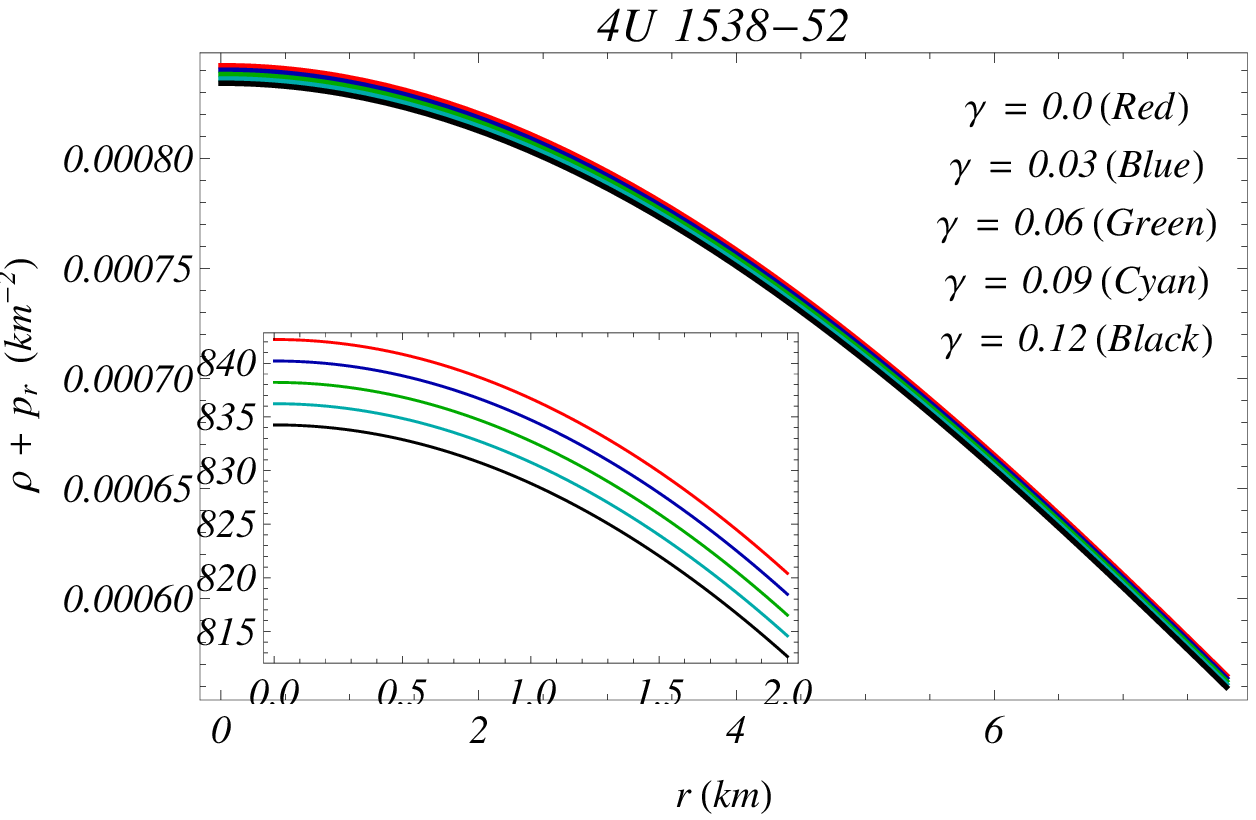}
        \includegraphics[scale=.45]{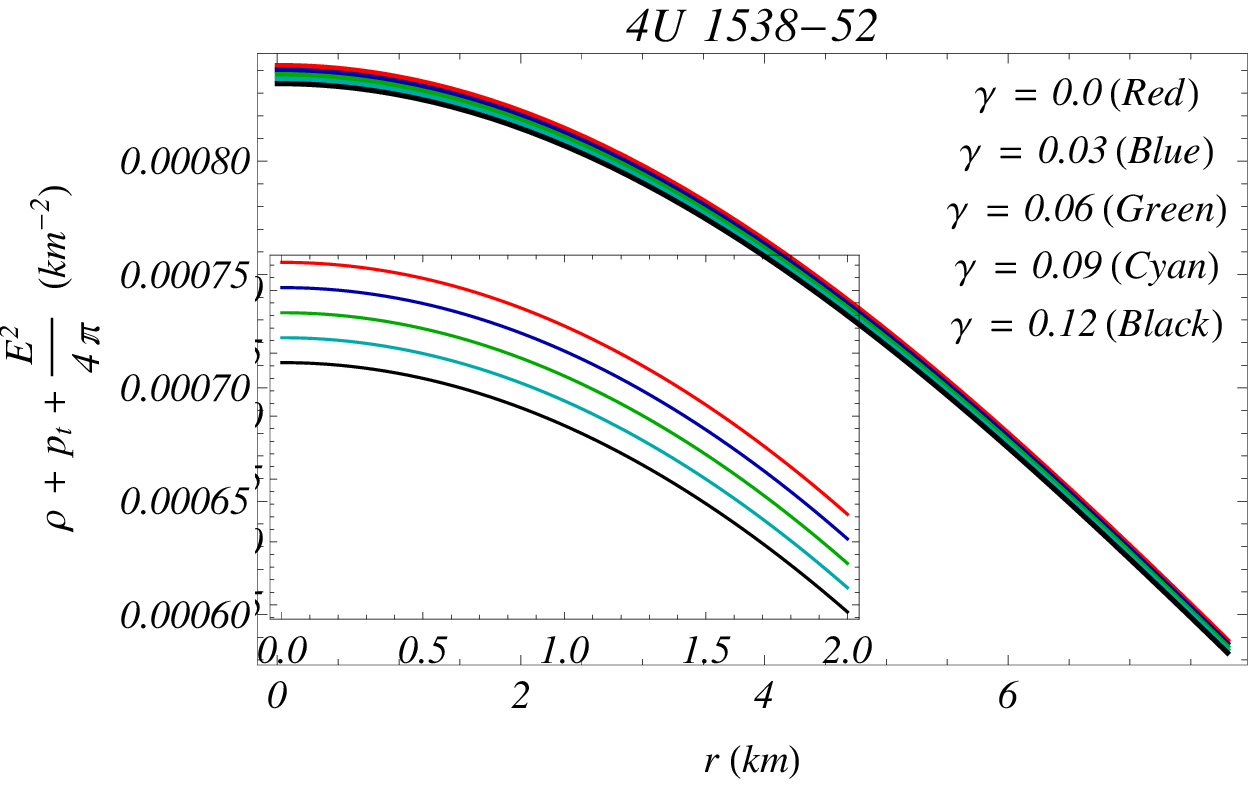}
        \includegraphics[scale=.45]{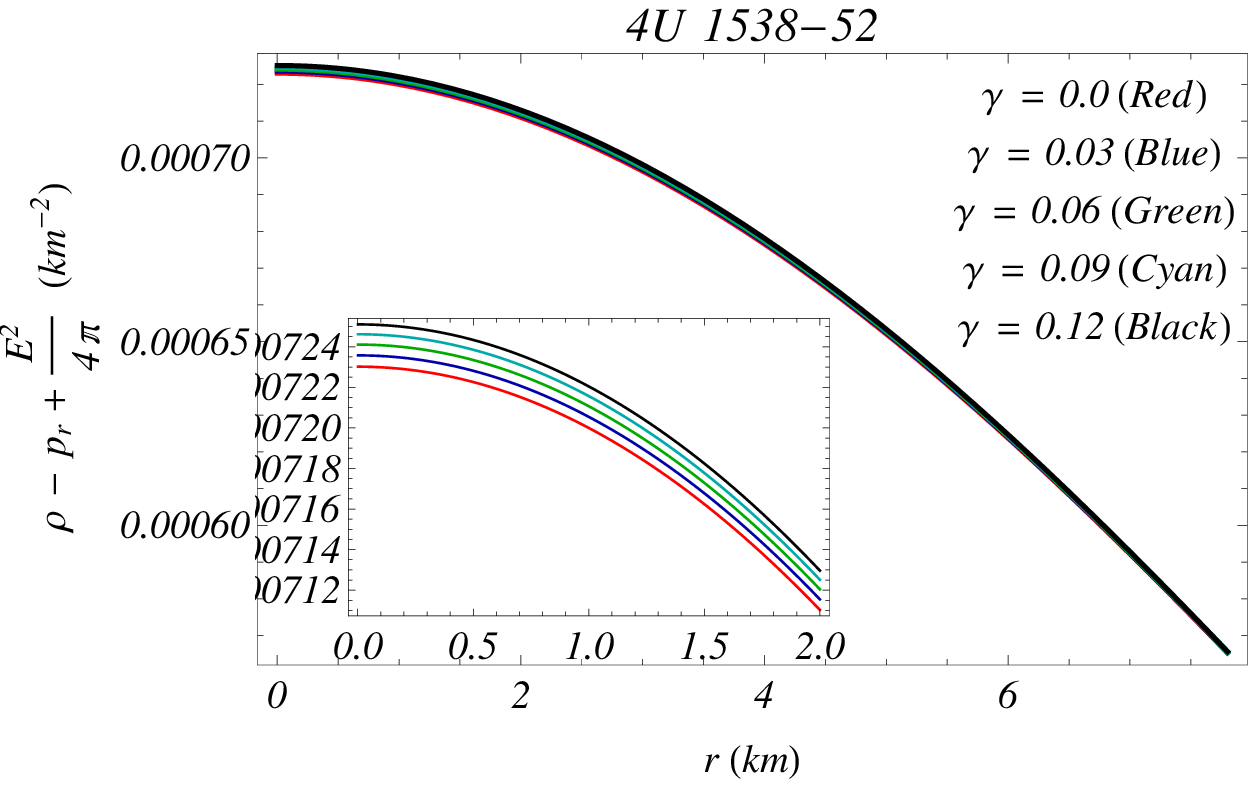}
       \caption{All the energy conditions are plotted inside the stellar interior for the strange star $4U 1538-52$ for different values of $\gamma$ mentioned in the figure.\label{ec}}
\end{figure}

The NEC is a minimum
requirement from SEC and WEC, i.e., if NEC is violated then both SEC and WEC will not be satisfied. The violation of these inequalities ensures the presence of exotic matter which occurred to described the model of wormhole in the context of the Einstein's general theory of relativity \cite{p101}. The existence of ordinary matter is
confirmed, if these conditions are satisfied. For the proposed star models, the validity of these conditions is checked
graphically in fig. \ref{ec}. It is found that our charged anisotropic model in $f(R,T)$ gravity satisfies all the above mentioned energy conditions for different values of $\gamma$ which ensures the presence of the ordinary matter inside the compact stars.

\subsection{ Equation of state parameter }
 The equation of
state parameters $\omega_r$ and $\omega_t$ are dimensionless quantity which describes the relation between matter density and pressures and these also represent the state of matter under a given set of physical conditions. When
$\omega_r,\,\omega_t$ lie between $0$ and $1$, it corresponds to the radiation era \cite{sharif187}. Using eqs. (\ref{l1})-(\ref{l3}) the equation of state parameters $\omega_r$ and $\omega_t$ for our present model are obtained as,
\[\omega_r=\frac{p_r}{\rho},~~\omega_t=\frac{p_t}{\rho}.\]
\begin{figure}[htbp]
    \centering
        \includegraphics[scale=.45]{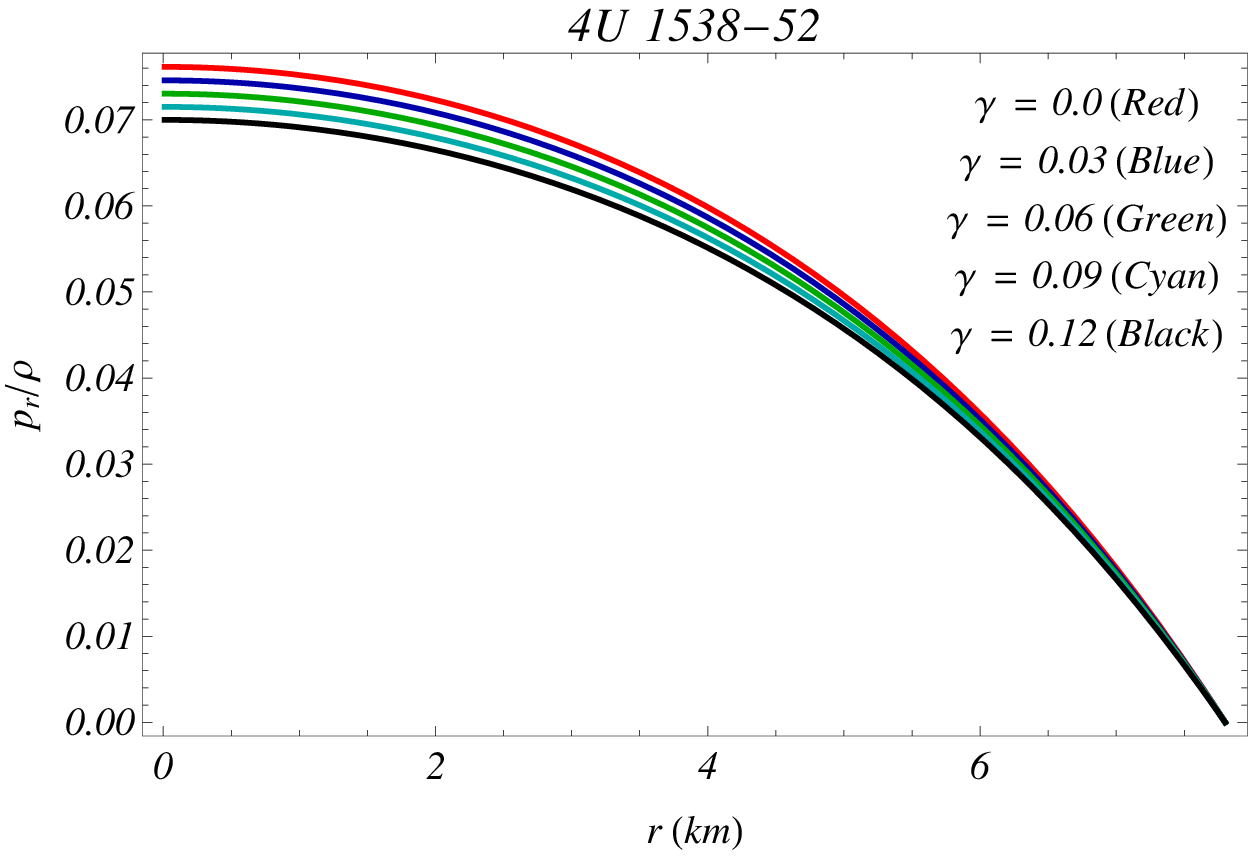}
        \includegraphics[scale=.45]{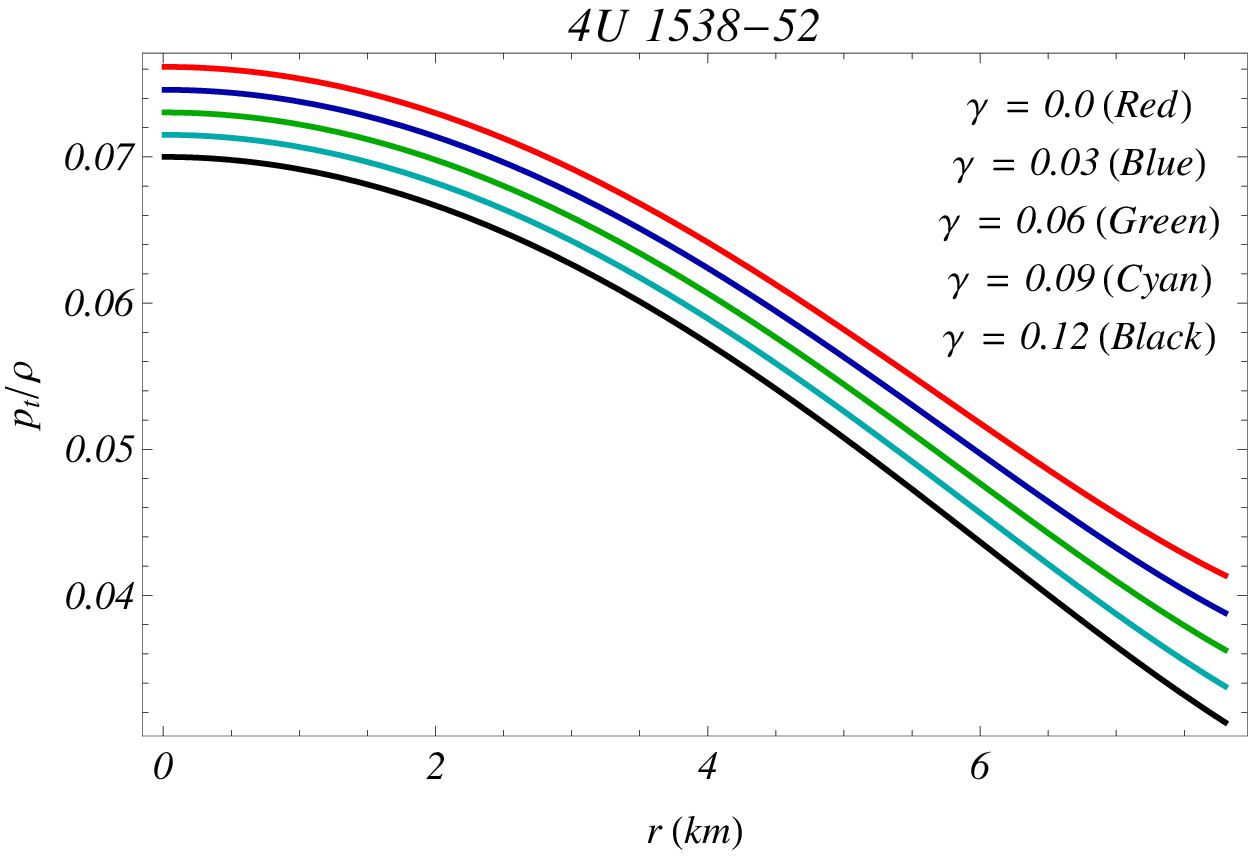}
       \caption{$p_r/\rho$ and $p_t/ \rho$ are shown against radius for different values of $\gamma$ mentioned in the figure.\label{eos} \label{omega}}
\end{figure}
The profiles of both $\omega_r,\,\omega_t$ are plotted in fig.~\ref{omega}. Both the profiles are monotonic decreasing function of r and also lies in the range $0<\omega_r,\,\omega_t<1$. So our present model of compact star in $f(R,T)$ gravity describes the radiating nature. It is also important to find out a relationship between the pressure and density which is known as the equation of state. To develop our model, we have assumed a non linear equation of state between the radial pressure and matter density, but, still we have no information about the relationship between the transverse pressure and the matter density. The behavior of radial and transverse pressure with respect to the matter density are shown graphically in fig.~\ref{eos}.


\begin{figure}[htbp]
    \centering
        \includegraphics[scale=.45]{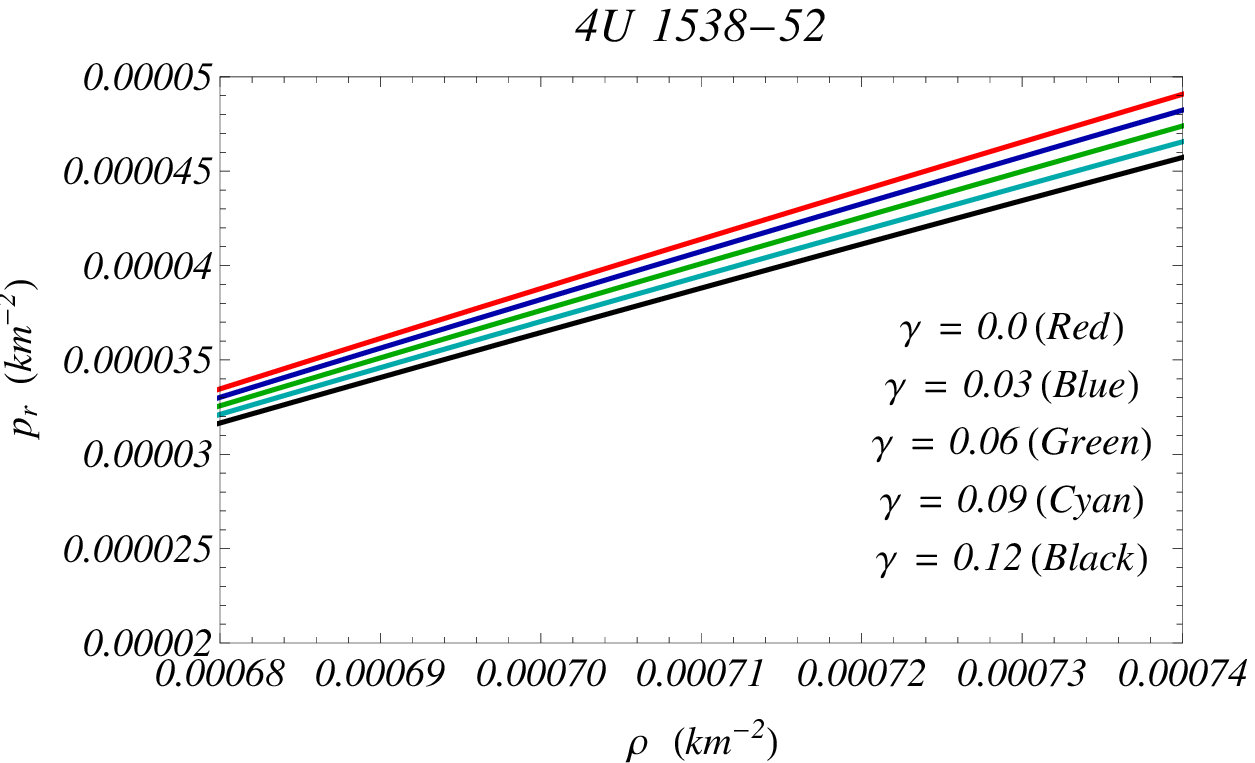}
        \includegraphics[scale=.45]{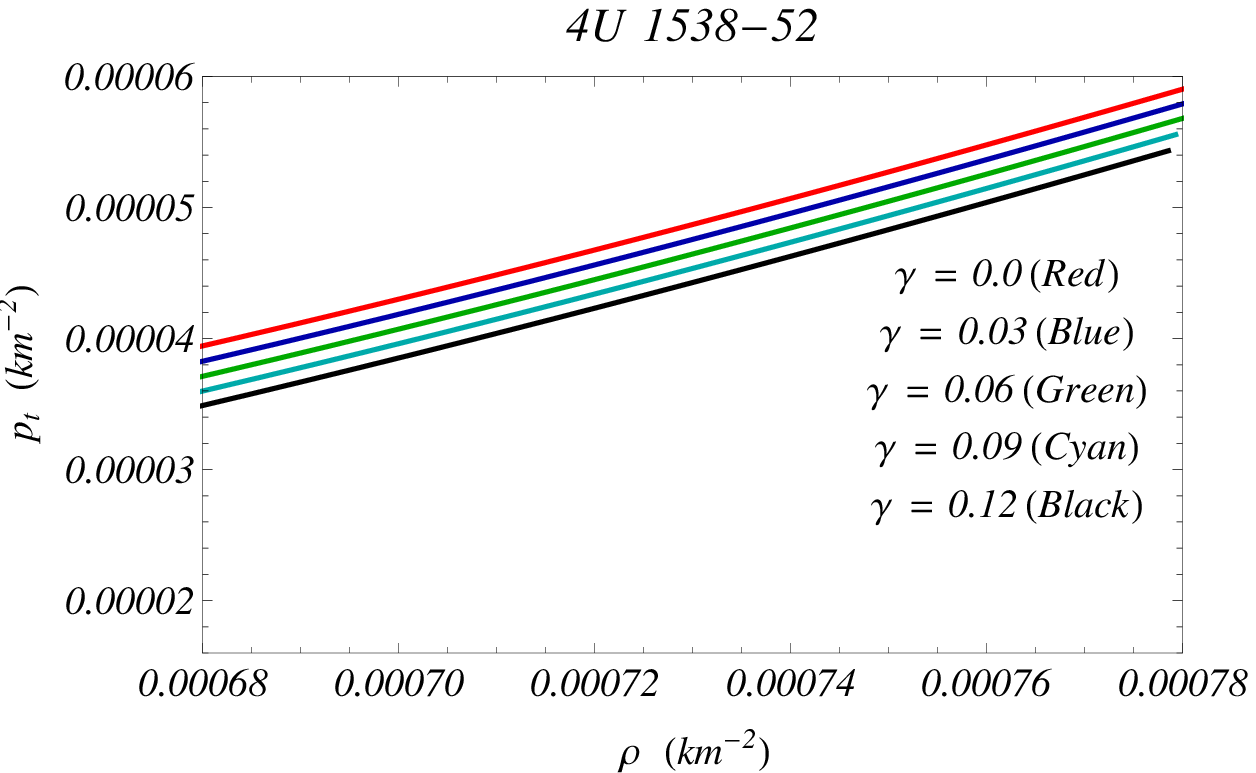}
       \caption{The variation of radial and transverse pressures are shown against matter density $\rho$ for different values of $\gamma$ mentioned in the figure.\label{eos}}
\end{figure}

\section{Mass radius relationship and surface redshift}

\begin{figure}[htbp]
    \centering
        \includegraphics[scale=.45]{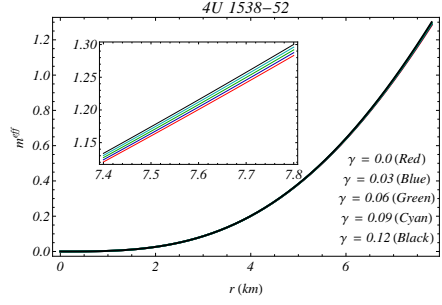}
       \caption{The variation of effective mass function is shown against radius for different values of $\gamma$ mentioned in the figure.\label{mass}}
\end{figure}

In our proposed charged model, the gravitational effective mass within the radius `r' can be obtained from the following formula \cite{murad}
\begin{eqnarray}\label{m4}
m^{\text{eff}}&=&4\pi \int_0^r \rho^{\text{eff}}(\tilde{r})~\tilde{r}^2 d\tilde{r} +\frac{q^2}{2r}+\frac{1}{2}\int_0^r\frac{q(\tilde{r})^2}{\tilde{r}^2}d\tilde{r},\nonumber\\
&=& m +\frac{\gamma}{2}\int_0^r(\rho-p_r-2p_t)(\tilde{r})~\tilde{r}^2 d\tilde{r} +\frac{q^2}{2r}+\frac{1}{2}\int_0^r\frac{q(\tilde{r})^2}{\tilde{r}^2}d\tilde{r}.
\end{eqnarray}
Where $m=4\pi \int_0^r \rho (\tilde{r})\tilde{r}^2 d\tilde{r}$, from equation (\ref{m4}), it is clear that for $\gamma=0$ both $m^{\text{eff}}$ and $m$ coincides. However by performing the above integration the effective mass function inside the radius `r' of the charged fluid sphere can be obtained as,
\begin{eqnarray}
m^{\text{eff}}&=&\frac{r}{2} \bigg[1 - e^{-A r^2}-\frac{e^{-A r^2}}{(1 + \alpha) (3 \gamma + 4 \pi)}\Big[2 e^{A r^2} \gamma^2 r^2 g_2(r)-\gamma \Big\{2 \alpha (-1 + e^{A r^2}) + (4 A \alpha + B + 3 \alpha B) r^2\nonumber\\&& -
 12 e^{A r^2} \pi r^2 g_2(r)+(1 + \alpha) B (-A + B) r^4 + 2 (-1 + e^{A r^2} + A r^2)\Big\}+4 \pi \Big\{1 + (-A + B) r^2\nonumber\\&& - \alpha (-1 + e^{A r^2} + 2 A r^2)+e^{A r^2}\left(-1+4\pi r^2 g_2(r)\right)\Big\}\Big]\bigg].
\end{eqnarray}
The effective mass function is regular at the center as $m^{\text{eff}}\rightarrow 0$ as $r\rightarrow 0$. The profile of the effective mass function is plotted in fig.~\ref{mass}.\par

The compactification factor inside the radius r for our present model is obtained as,
\begin{eqnarray}
u^{\text{eff}}=\frac{m^{\text{eff}}}{r},
\end{eqnarray}
As adressed by Giuliani and Rothman \cite{giu}, the problem of
finding a lower bound on the radius $R$ of a charged sphere with mass
M and total charge $Q$ is given by, $Q < M $ and in this case, collapse always takes place
at a critical radius $R_c$ outside the outer horizon, and as $Q\rightarrow M$,
this value approaches the horizon. The upper bound of the mass of charged sphere
was generalized by Andr\'{e}asson \cite{an1} as
\begin{eqnarray}
\sqrt{M} \leq \frac{\sqrt{R}}{3}+\sqrt{\frac{R}{9}+\frac{Q^2}{3R}},\label{w1}
\end{eqnarray}
by assuming the inequality $\rho-p_r-2p_t \geq 0.$
The equation (\ref{w1}) equivalently gives,
\begin{eqnarray}
\frac{M}{R}\leq \left(\frac{1}{3}+\sqrt{\frac{1}{9}+\frac{Q^2}{3R^2}}\right)^2,\label{w2}
\end{eqnarray}
One can easily check that the eqn. (\ref{w2}) obeys the Buchdahl's limit \cite{buch} $\frac{2M}{R}<\frac{8}{9}$ for uncharged case.\par
On the other hand, B\"{o}hmer
and Harko \cite{harko15} proposed the lower bound
of mass to the radius for charged fluid sphere as,
\begin{eqnarray}
\frac{3Q^2}{4R^2}\frac{1+\frac{Q^2}{18R^2}}{1+\frac{Q^2}{12R^2}} \leq \frac{M}{R},\label{w3}
\end{eqnarray}
combining (\ref{w2}) and (\ref{w3}) we get,
\begin{eqnarray}
\frac{3Q^2}{4R^2}\frac{1+\frac{Q^2}{18R^2}}{1+\frac{Q^2}{12R^2}} \leq \frac{M}{R} \leq \left(\frac{1}{3}+\sqrt{\frac{1}{9}+\frac{Q^2}{3R^2}}\right)^2.\label{w4}
\end{eqnarray}
In eqn.~(\ref{w4}), $R$ represents the radius of the fluid distribution, $m^{\text{eff}}(r = R) = M,\,q(r=R)=Q$, $M$ and $Q$ are respectively the gravitational mass and total charge inside the fluid sphere.

\begin{center}
\begin{tabular}{ c | c |c  |c }
$\gamma$ &  Value of lower & $u^{\text{eff}}(R)$&  Value of upper\\
&limit of eq.~(\ref{w4})& &limit of eq.~(\ref{w4})\\
\hline
$0.0$ &$0.00749792$&$0.164519$&$0.666717$\\
$0.03$ &$ 0.0260527$&$0.165074$&$0.667271$\\
$0.06$ &$0.035939$&$0.16562$&$0.667816$ \\
$0.09$ &$0.0435352$&$0.166159$&$0.668353$ \\
$0.12$&$0.0499014$&$0.166691$&$0.668881$\\
\hline
\end{tabular}
\end{center}

From the above table we see that the inequality given in eqn.~\ref{w4} is satisfied by our present model for different values of $\gamma$.

\begin{table*}[t]
\centering
\caption{The values of the constants $a,\,C$ and $D$ for the compact star 4U 1538-52 by assuming $M = 0.87~M_{\odot},\, R = 7.8 $ km., Q = 0.078.}
\begin{tabular}{@{}cccccccccccccccc@{}}
\hline
$\gamma$&$\alpha$&$\beta$&$\rho_c$&$\rho_s$&$p_c$&$\Gamma_{r0}$ & $M^{\text{eff}}$ & $z_s^{\text{eff}}(R)$\\
\hline
0.0& 0.15916& $5.08405 \times 10^{-8}$& $1.05601 \times 10^{15}$& $7.62612  \times 10^{14}$ & $ 7.23777 \times 10^{34}$ & 3.4221& 1.28325 & 0.220819\\
0.03& 0.15539& $4.93998 \times 10^{-8}$& $ 1.05503      \times 10^{15}$ &  $7.60796 \times 10^{14}$ & $ 7.08216  \times 10^{34}$ & 3.4029&1.28758 &0.221829 \\
0.06&0.151689& $4.79945 \times 10^{-8}$& $ 1.05404     \times 10^{15}$&  $ 7.58988 \times 10^{14}$ & $ 6.9286   \times 10^{34}$ & 3.38409 & 1.29184&0.222827\\
0.09&0.148057& $4.66236\times 10^{-8}$&  $1.05305   \times 10^{15}$&  $7.57189 \times 10^{14}$ & $  6.77705 \times 10^{34}$ &3.36564 & 1.29604& 0.223814\\
0.12&0.144492& $4.5286\times 10^{-8}$& $1.05204  \times 10^{15}$& $ 7.55399  \times 10^{14}$& $  6.62747 \times 10^{34}$ & 3.34757 & 1.30019 & 0.224789\\
\hline
\end{tabular}
\end{table*}\label{t101}

\begin{figure}[htbp]
    \centering
        \includegraphics[scale=.45]{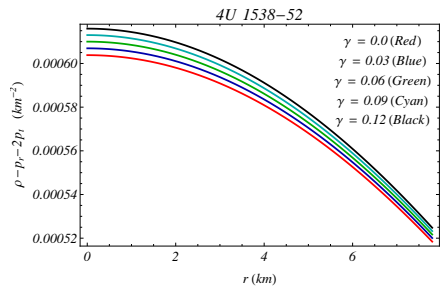}
       \caption{The variation of $\rho-p_r-2p_t$ with respect to radius is shown for different values of $\gamma$ for the compact star 4U1538-52.\label{diag}}
\end{figure}

\begin{figure}[htbp]
    \centering
        \includegraphics[scale=.45]{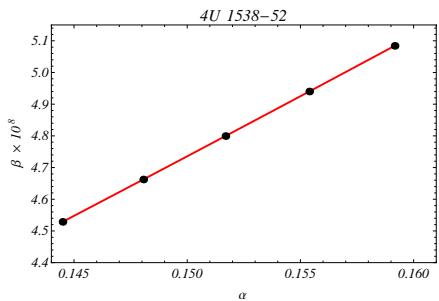}
       \caption{The variation of $\beta$ with respect to $\alpha$ has been depicted.\label{par}}
\end{figure}

\section{Discussion}

In the
present work,
in the context of modified theory of gravity, we have obtained a new model of charged compact star. To explore the model we have considered the functional forms of $f(R, T)$ as $f(R,T)=R +2 \gamma T$ where $R$ and $T$ are respectively the Ricci scalar and the trace of energy-momentum tensor $T_{\mu \nu}$ respectively. For our present analysis, we have choose $\gamma$ as a small positive constant since for negative values of $\gamma$ produces negative value of the electric field $E^2$, we have exclude this case by considering the physical acceptability of the present model. The model has been developed by taking Krori-Barua (KB) {\em ansatz} since it is well known that KB metric produces a singularity free model. All numerical calculations and plots have been done for the strange star candidate 4U 1538-52 whose observed mass and radius are given by $(0.87 \pm 0.07)~M_{\odot}$ and $7.866 \pm 0.21$ km. respectively. \par

To obtain the result in closed form, instead of choosing adhoc expression for electric field $E^2$, we have chosen non linear equation of state as Chaplygin form: $p_r=\alpha \rho -\frac{\beta}{\rho}$. Here $\alpha$ and $\beta$ are small positive constants. From our analysis we have shown that the values of $\alpha$ and $\beta$ depend on the coupling constant $\gamma$ and their numerical values have been obtained for the compact star 4U 1538-52 in table~I. It is clear from the table that both $\alpha$ and $\beta$ decreases if $\gamma$ increases and it is to be noted that the effect of $\beta$ is very small to the model compare to alpha. The variation of $\beta$ with respect to $\alpha$ is shown in fig.~\ref{par}. We have also obtained the numerical values of the central density, surface density and central pressure in the order of $10^{15}$~gm.cm$^{-3}$, $10^{14}$~gm.cm$^{-3}$ and $10^{34}$~dyne.cm$^{-2}$ respectively. The numerical values of the central density, surface density and central pressure all decreases with the increasing value of $\gamma$. On the contrary the numerical values of the effective mass and surface redshift increases as $\gamma$ increases. The present model is potentially stable as well as the causality condition is satisfied. All the energy conditions are verified for our model with the help of graphical representation. The effect of coupling parameter $\gamma$ on the different physical parameters like, density, pressure, anisotropic factor, sound velocity, compactness factor, mass function have been widely discussed.

The surface stress energy ($\sigma$) and surface pressure ($\mathcal{P}$) for our present model are obtained as,
\begin{eqnarray*}
  \sigma &=& -\frac{1}{4\pi R}\Big[\sqrt{1-\frac{2M}{R}+\frac{Q^2}{R^2}}-e^{-\frac{aR^2}{2}}\Big], \\
  \mathcal{P}&=& \frac{1}{8\pi R}\left[\frac{1-\frac{M}{R}}{\sqrt{1-\frac{2M}{R}+\frac{Q^2}{R^2}}}-(1+BR^2)e^{-\frac{aR^2}{2}}\right].
\end{eqnarray*}

From our entire analysis, we can conclude that, the present model of compact star, more or less, behaves like GR.

{\bf ACKNOWLEDGMENTS:} PB is thankful to IUCAA, government of India for providing visiting associateship.

\end{document}